\documentclass[aps,prb,reprint,preprintnumbers]{revtex4-1}
\usepackage{graphicx}
\usepackage{bm}
\usepackage{epsfig}
\usepackage{ulem}
\usepackage{color}
\usepackage{mathrsfs}
\usepackage{dcolumn}
\usepackage{setspace}
\usepackage{array}
\usepackage{amsmath}
\usepackage{amssymb}
\usepackage{dsfont}
\usepackage{gensymb}
\usepackage{dcolumn}
\usepackage{multirow}
\usepackage{bibentry,natbib}
\usepackage{booktabs}
\begin{document}
\title{Geometric Induction in Chiral Superfluids}

\author{Qing-Dong Jiang$^1$$^2$}
\email{qingdong.jiang@sjtu.edu.cn}
\author{A. Balatsky$^3$$^4$}
\affiliation{{}\\ $^1$ Tsung-Dao Lee Institute, Shanghai Jiao Tong University, Shanghai 200240, China\\
$^2$School of Physics and Astronomy, Shanghai Jiao Tong University, Shanghai 200240, China\\
$^3$Nordita, KTH Royal Institute of Technology and Stockholm University, Roslagstullbacken 23, SE-106 91 Stockholm, Sweden\\
$^4$ UCONN, Department of Physics, Storrs, CT 06269, USA}
\begin{abstract}
We explore the properties of chiral superfluid thin films coating a curved surface. Due to the vector nature of the order parameter, a geometric gauge field emerges and leads to a number of observable effects such as anomalous vortex-geometric interaction and curvature-induced mass/spin supercurrents. We apply our theory to several well-known phases of chiral superfluid $\rm ^3 He$ and derive experimentally observable signatures. We further discuss the cases of flexible geometries where a soft surface can adapt itself to compensate for the strain from the chiral superfluid. The proposed interplay between geometry and chiral superfluid order provides a fascinating avenue to control and manipulate quantum states with strain. 
\end{abstract}
\maketitle


Geometric phases, rooted in the concept of parallel transport and related to topology, figure prominently in a startling variety of physical contexts, ranging from optics and hydrodynamics to quantum field theory and condensed matter physics \cite{ShapereW}.
In classical systems, for example, the geometric phase shift of the Foucault pendulum is equal to the enclosed solid angle subtended at the earth's center \cite{foucault}. Other classical examples of geometric phases include the motion of deformable bodies \cite{ShapereW2} and tangent-plane order on a curved substrate \cite{ParkL,ATuner}. In quantum mechanics, the geometric phases arise from slowly transporting an eigenstate round a circuit C by varying parameters $\bold R$ in its Hamiltonian $\hat {\bold H}(\bold R)$ \cite{VinitskiiD}. For example, the geometric phase of a single-electron Bloch wave function in the Brillouin zone is essential for topological states of matter such as the quantum Hall effect and topological insulators \cite{XiaoCN}.

Beyond the single-electron picture, the concept of geometric phase has become a defining property of topological superconductors, where Cooper pairs can directly inherit their geometric phases from the two paired electrons \cite{QiZ}.
Chiral superconductors, a particularly interesting class of topological superconductors \cite{Maeno}, have received great attention due to their promise of hosting Majorana zero modes in vortex cores and at edges, which are central to several proposals for topological quantum computation \cite{Kopnin,Volovik}.

In a chiral {\it $p$}-wave superconductor, the Cooper pairs carry orbital angular momentum (OAM) of $\hbar$, and the order parameter is a complex vector defined in the tangent plane of a two-dimensional (2D) surface $|\Psi\rangle=\psi~\left(\bold{\hat e_1}\pm i\,\bold{\hat e_2}\right)/\sqrt{2}$
with $\bold{\hat e_1}$ and $\bold{\hat e_2}$ the local orthogonal basis and $\psi$ the complex amplitude \cite{Volovik,JSauls,Volovik1}. Here $\pm$ sign denotes the chirality and the direction of the OAM.  When such an order parameter with positive chirality evolves in a circuit on a curved 2D surface (Fig. 1 as an illustration), a geometric phase arises according to the formula $\frac{1}{\langle \Psi|\Psi\rangle}\oint_C {\langle \Psi| i\partial_\mu|\Psi\rangle} dl^\mu=\oint_C  \omega_\mu dl^\mu $. Here $\omega_\mu=\mathbf{\hat e_1}\cdot \partial_\mu \mathbf{\hat e_2}$ is the geometric connection whose curl is Gaussian curvature (see Sec. S-I in the Supplemental Material, where we present the mathematical foundation of geometric connection \cite{Append}).  Generalization to a chiral $\ell$-wave order parameter, describing a condensate of Cooper pairs with orbital angular momentum $\ell \hbar$, yields a geometric phase $\ell \oint_C  \omega_\mu dl^\mu$ \cite{Append}.
\begin{figure}[!htb]
\includegraphics[height=3cm, width=8cm, angle=0]{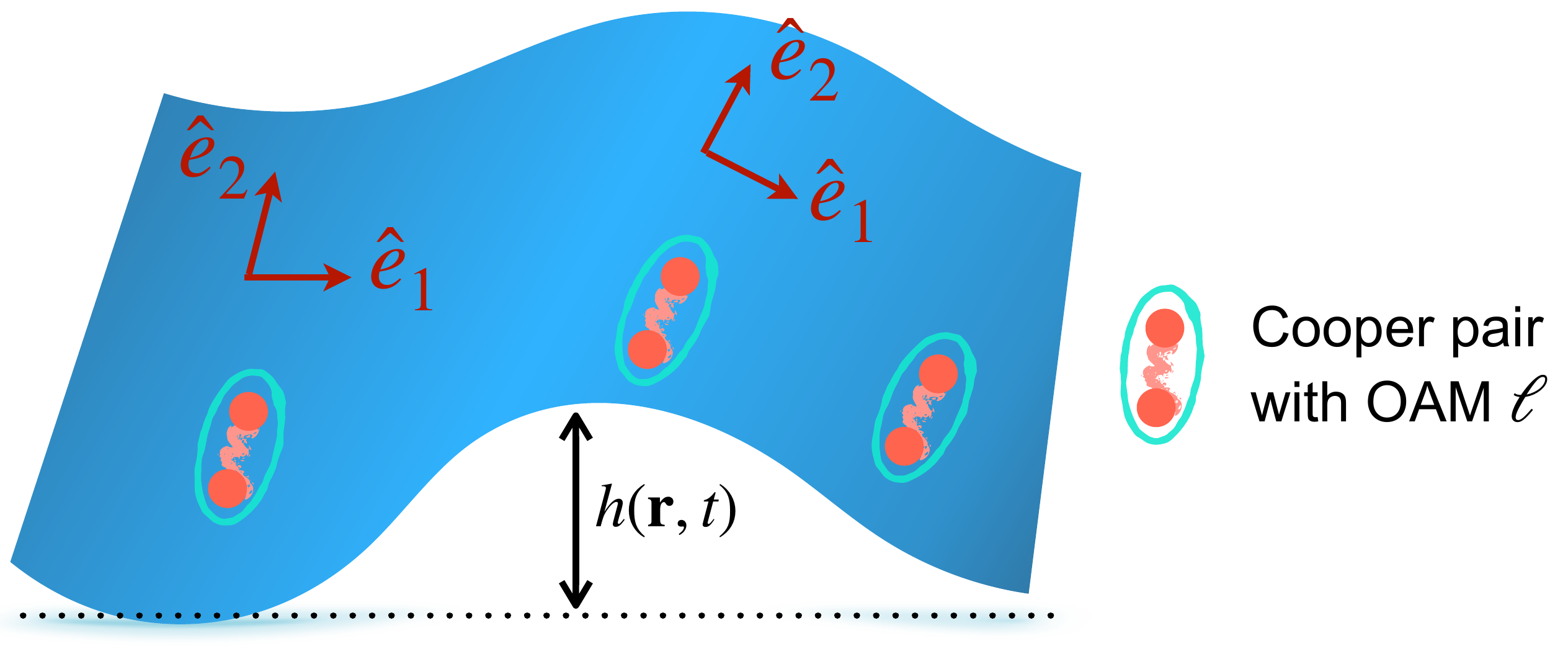}
\caption{Schematic illustration of transporting a vectorial order on a curved surface.
The height $h(\mathbf r, t)$ measures the deviation of a curved surface from a plane. \label{fig1}}
\end{figure}
The geometric connection $\omega_\mu$ may lead to a number of intriguing effects, such as the geo-Meissner effect \cite{KvorningH} and the geometric Josephson effect \cite{JiangHW}, which serve as definitive signatures of chiral superconductivity.

In this Letter, we study the interplay between chiral superfluidity and geometry.  We are motivated by the following observations: i) Chiral superfluids are charge-neutral condensates. Therefore, the corresponding electromagnetic signature must be qualitatively different from that of superconductors. ii) Unlike chiral superconductors, chiral superfluids are observed in nature ($\rm ^3$He-A phase) \cite{Vollhardt} and provide a testbed for our proposed geometric induction theory. iii) The study of interactions between chiral-superfluid vortices and geometry, while experimentally feasible, is still lacking in the literature. iv) Geometry may provide a practical knob to manipulate novel quantum states, such as the Majorana zero mode in a vortex. Thus it may offer a unique route to quantum manipulation including braiding - central to topological quantum computation \cite{Nayak,chung}.

The paper is organized as follows: We first develop the necessary formalism for 2D chiral superfluids covering a curved surface. We then study the interaction between vortices and geometry, aiming at controlling quantum states with geometry. Next, we derive mass current and spin current induced by Gaussian curvature in several well-known phases of chiral superfluid $^3$He, and we obtain the associated electromagnetic signatures. Finally, we study the quantum backaction of a chiral superfluid on a flexible surface.

\textit{\bf Emergent geometric gauge fields.---} The order parameter of a chiral $\ell$-wave superfluid can be generically written as a rank-$\ell$ tensor, i.e.,
\begin{equation}\label{OP}
\Psi=\psi~\underbrace{\epsilon_\pm\otimes\epsilon_\pm\dots\otimes \epsilon_\pm}_{\text{$\ell$ times}},
\end{equation}
where  $\epsilon_{\pm}=\frac{1}{\sqrt{2}}\left(\mathbf{\hat e_1}\pm i\,\mathbf{\hat e_2}\right)$ denote chiral basis, and $\psi=\sqrt{\rho}e^{i\theta}$ is the complex amplitude in terms of the superfluid density $\rho$ and phase $\theta$. $\ell=1$ ($\ell=2...$) corresponds to the order parameter of chiral {\it p-wave} ({\it d-wave}...) superfluids. In this paper, we consider the positive chirality. The negative chirality cases can be obtained by reversing the sign of $\ell$ in our formulas.

On a curved surface (substrate), the minimal Lagrangian of a chiral $\ell$-wave superfluid reads
\begin{equation}
\label{eq2}
\mathcal L_{\text{sf}}=i\hbar\psi^*D_t\psi-\frac{\hbar^2 g^{ij}}{2m}(D_i \psi)^*(D_j \psi)-V(|\psi|),
\end{equation}
where $g^{ij}$ is inverse the metric tensor $g_{ij}$, $m$ is the mass of a Cooper pair, and $V(|\psi|)$ is a symmetry-breaking potential. Since the order parameter $\psi(t,\bold r)$ depends on the choice of orthonormal basis $\mathbf{\hat e_1}$ and $\mathbf{\hat e_2}$, one needs to use the covariant derivatives $D_\mu$ $(\mu=0,1,2)$ defined by
\begin{equation}\label{eq3}
D_\mu=\partial_\mu+i\ell \,\omega_\mu,
\end{equation}
where $\omega_\mu=\mathbf{\hat e_1}\cdot \mathbf{\partial_\mu\hat e_2}$ is the geometric connection originating from parallel transport of a vector on a curved surface \cite{Append}. The geometric connection $\omega_\mu$ is a geometric gauge field akin to the electromagnetic vector potential, with the Gaussian curvature playing the role of a magnetic field. It was shown that a similiar Lagrangian can induce Hall viscosity \cite{Hoyos} and thermal Hall effect \cite{GolanS}. From the covariant derivatives, we can obtain the total field strength tensor
$T_{\mu\nu }=i\,\left[D_\mu,D_\nu \right]=-\ell \,G_{\mu\nu },$
where $G_{\mu\nu }=\partial_\mu \omega_\nu  -\partial_\nu  \omega_\mu$ is the geometric field tensor, and correspondingly, we define the electric- and magneticlike field strength:
\begin{equation}\label{GInduction}
{\mathcal E}^i=\frac{1}{2}\frac{\epsilon^{i\mu\nu }}{\sqrt{g}} G_{\mu\nu },~~ \mathcal B=\frac{1}{2}\frac{\epsilon^{0ij}}{\sqrt{g}}G_{ij}.
\end{equation}
with $i,j$ taking values $1$ or $2$.
Physically, $\mathcal B$ is the Gaussian curvature of a curved surface, and the meaning of $\boldsymbol{\mathcal E}$ will be clear later. In what fellows, we will discuss a number of effects that originate from the geometric gauge field.

{\bf Anomalous vortex-geometry interaction.---}
To discuss vortex physics, we rewrite Eq.\eqref{eq2} in terms of superfluid density $\rho$ and phase $\theta$, i.e., set $\Psi=\sqrt{\rho}e^{i\theta}$ to get
\begin{eqnarray}
\mathcal L_{\text{sf}}&=&i\hbar \rho \left(\partial_0\theta+\ell \omega_0\right)-\frac{\hbar^2 \rho g^{ij}}{2m}(\partial_i \theta + \ell \omega_i)(\partial_j \theta+ \ell \omega_j)\nonumber\\
&&-V(\rho),
\end{eqnarray}
where the potential $V(\rho)=A\left(\rho-\bar \rho\right)^2$ guarantees that the superfluid acquires a finite average density $\bar\rho$. Upon integrating out the fluctuations of density, one obtains
\begin{eqnarray}
\mathcal L_{\text{sf}}&=&\frac{\gamma_0}{2} \left(\partial_0\theta+ \ell \,\omega_0\right)^2-\frac{\gamma_s}{2}(\boldsymbol\nabla \theta  + \ell \,\boldsymbol \omega)^2,
\end{eqnarray}
where $\gamma_0={\hbar^2}/{2A}$ indicates fluctuation strength and $\gamma_s={\hbar^2 {\bar\rho}}/{m}$ denotes the superfluid stiffness.
Upon rescaling temporal and spatial coordinates, we arrive at an effective Lagrangian density of the Lorentz-invariant form:
\begin{equation}\label{LorInvLag}
\mathcal L_{\text{eff}}=\frac{\gamma}{2}\left(\partial_\mu \theta+\ell\,\omega_\mu\right)^2.
\end{equation}

To discuss vortex interactions and dynamics, we introduce the alternative form
\begin{equation}\label{AuxLorInvLag}
\mathcal L_{\text{eff}}=-\frac{1}{2\gamma}\xi_\mu^2+\xi^\mu\left(\partial_\mu \theta+\ell\,\omega_\mu\right),
\end{equation}
which gives Eq.\eqref{LorInvLag} after integrating out the auxiliary field $\xi^\mu$.
Without loss of generality, one can take the phase $\theta$ as a smoothly fluctuating field, except that at vortices where it winds around $2\pi$ \cite{Zee}. Therefore, one can write $\partial_\mu \theta=\partial_\mu \theta_{\text{smooth}}+\partial_\mu \theta_{\text{vortex}}$, and plug it into the Eq.\eqref{AuxLorInvLag}, yielding
\begin{equation}
\mathcal L_{\text{eff}}=-\frac{1}{2\gamma}\xi_\mu^2+\xi^\mu\left(\partial_\mu \theta_{\text{smooth}}+\partial_\mu \theta_{\text{vortex}}+\ell\, \omega_\mu\right).
\end{equation}
Integrating out $\theta_{\text{smooth}}$, we get the constraint for $\partial_\mu \xi^\mu=0$, which can be automatically satisfied by the substitution $\xi^\mu\equiv\varepsilon^{\mu\nu \lambda}\partial_\nu a_{\lambda}$. Notice that, on a curved surface, $\varepsilon^{\mu\nu \lambda}\equiv \epsilon^{\mu\nu \lambda}/\sqrt{g}$, and  $a_\mu$ can be understood as a gauge field, because the change $a_\mu\rightarrow a_\mu+\partial_\mu \Gamma$ does not change $\xi^\mu$. With this substitution, we can write the action in terms of $a_\mu$:
\begin{eqnarray}\label{ELV}
\mathcal S_{\text{eff}}=\int dtd^2r\sqrt{g}\left[ -\frac{f_{\mu\nu}^2}{4\gamma} +a_\lambda \varepsilon^{\lambda\nu\mu}\partial_\nu \left(\partial_\mu \theta_{\text{vortex}}+\ell\,\omega_\mu\right)\right]\nonumber
\end{eqnarray}
where $f_{\mu\nu }=\partial_\mu a_\nu -\partial_\nu  a_\mu$ is the strength tensor of the $a_\mu$ field.
We reveal the physical meaning of the second term of the above equation. Integrating the zero component $\varepsilon^{0\mu\nu }\partial_\nu  \partial_\mu \theta_{\text{vortex}}$ over a region containing a vortex yields $\int d^2r\sqrt{g}\, \varepsilon^{0\mu\nu}\partial_\mu \partial_\nu \theta_{\text{vortex}} =\oint d\bold r\cdot \boldsymbol{\nabla} \theta_{\text{vortex}}=2\pi$. We thus recognize  $\varepsilon^{0\mu\nu }\partial_\mu \partial_\nu \theta_{\text{vortex}}$ as the density of vortices, i.e., the time component of a vortex current density
\begin{eqnarray}
j_{\text{vor}}^\lambda&=&\varepsilon^{\lambda\mu\nu}\partial_\mu \partial_\nu  \theta_{\text{vortex}}.
\end{eqnarray}
One the other hand, we realize that $\varepsilon^{0\mu\nu}\partial_\mu \omega_\nu =\mathcal B$ and $\varepsilon^{i\mu\nu }\partial_\mu \omega_\nu=\mathcal E^i$ are the geometric field strength defined in Eq.\eqref{GInduction}.
Therefore, we also identify a geometric current
\begin{eqnarray}
j_{\text{geo}}^\lambda&=&\varepsilon^{\lambda\mu\nu}\partial_\mu \omega_\nu =\left(\mathcal B, \mathcal E^1, \mathcal E^2\right).
\end{eqnarray}
Substituting vortex current and geometric current into the effective action, we obtain the effective Lagrangian density for vortices and geometry
\begin{equation}\label{eq12}
\mathcal L_{\text{vor-geo}}=-\frac{1}{4\gamma} f_{\mu\nu}^2 + a_\lambda \left(j_{\text{vor}}^\lambda+j_{\text{geo}}^\lambda\right).
\end{equation}
This {\bf central equation} governs the dynamics and interactions of vortices and geometry in a chiral superfluid covering a curved surface. There are three types of interactions mediated by the gauge field $a_\mu$, namely {\it vortex-vortex} interaction, {\it geometry-geometry} interaction, and {\it vortex-geometry} interaction.
The vortex-geometry interaction resembles the quasiparticle-geometry coupling (the Wen-Zee term \cite{WenZee1992}) of quantum Hall (QH) liquids. In SM \cite{Append}, we derive an alternative form of Eq.\eqref{eq12}, revealing the similarity and differences between chiral superfluidity and QH physics. While the analogy has been realized in the literature previously \cite{StoneRoy2004,GolkarMoroz}, the field theory of chiral superfluidity has two key differences with QHE: the gauge field action is Maxwell-like instead of Chern-Simons action and we have Aharonov-Casher gauge potential term absent in QHE. These differences lead to qualitatively different electromagnetic responses.

In the static limit, Eq.\eqref{eq12} can be understood by analogy to the Coulomb gas model: the Gaussian curvature $\mathcal B(\bold r)$ plays the role of a non-uniform background charge distribution and the vortices appear as point-like sources with electrostatic charges equal to their winding number.  As a result, the vortices tend to position themselves so that the Gaussian curvature is screened: the negative ones on maximum or minimum while the positive ones on the saddles of a surface.
\begin{figure}[!htb]
\includegraphics[height=3.2cm, width=8.6cm, angle=0]{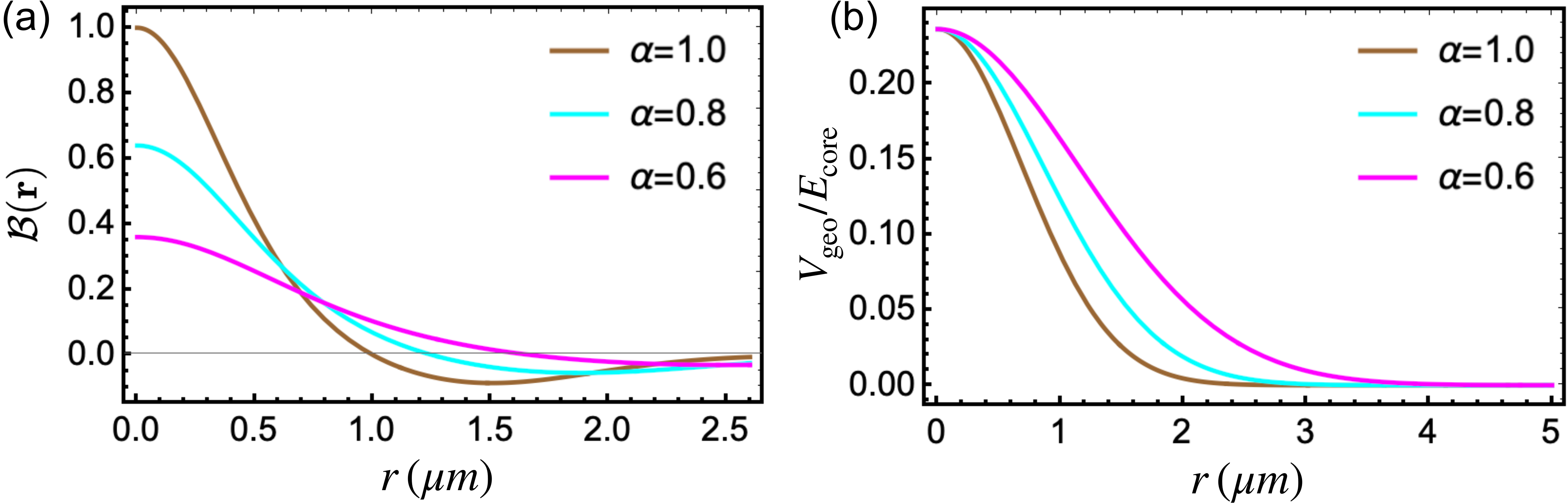}
\caption{(a) shows the spatial-dependent Gaussian curvature of a Gaussian bump of three different aspect ratios $\alpha=1,\, 0.8$ and $0.6$.  (b) shows the corresponding spatial-dependent geometric potential, and $E_{\rm core}\approx \hbar^2 \rho_s/m$ is a typical 2D vortex core energy\cite{Mmondal}. \label{fig2}}
\end{figure}

Let us quantify the strength of vortex-geometric interaction by considering a vortex in a rotational symmetric 2D surface specified by a three-dimensional vector $\bold R(r,\varphi)=\left(r\cos \varphi, r\sin\varphi,h_0 \exp{\left(-r^2/2r_0^2\right)}\right)$, where $r$ and $\varphi$ are plane polar coordinates. Clearly, $\bold R(r,\varphi)$ describes a static Gaussian bump with a maximum height $h_0$ and spatial extent $\sim r_0$. It is useful to characterize the deviation of the bump from a plane in terms of a dimensionless aspect ratio $\alpha\equiv h_0/r_0$. We can define local orthonormalized basis vectors $\mathbf{\hat  e_r}$ and $\boldsymbol{\hat e_\varphi}$ by normalizing two orthogonal tangent vectors $\bold t_r=\partial \bold R/\partial r$ and $\bold t_\varphi=\partial \bold R/\partial\varphi$. The components of the geometric gauge field introduced in Eq.\eqref{eq3} are given by $\omega_i=\mathbf{\hat e_r}\cdot\partial_i \boldsymbol{\hat e_\varphi}$, i.e., $\omega_r=0$ and $\omega_\varphi=-1/\sqrt{c(r)}$ with $c(r)\equiv 1+\frac{\alpha^2 r^2}{r_0^2}\exp \left(-\frac{r^2}{r_0^2}\right)$.
Consequently, the Gaussian curvature of the bump can be obtained $\mathcal B(r)=\frac{\alpha^2 }{r_0^2c(r)^2}\left(1-\frac{r^2}{r_0^2}\right)\exp{(-r^2/r_0^2)}$,
which generates a geometric potential
\begin{eqnarray}
V_{\text{geo}}(\bold r)=\int d^2 r^\prime\sqrt{g(\bold r^\prime)}\, \mathcal B(\bold r^\prime)\,\Gamma(\bold r,\bold r^\prime)
\end{eqnarray}
via the propagator $\Gamma(\bold r^\prime, \bold r)$ of the gauge field $a_\mu$. Here $g(\bold r^\prime)=c(r^\prime)$ is the determinant of the metric. One can employ a conformal transformation to obtain the propagator $\Gamma(\bold r^\prime, \bold r)$ and then the geometric potential  \cite{Append}
\begin{equation}\label{eqvgeo}
V_{\text{geo}}(r)={\frac{\hbar^2\rho_s}{m}}\int_r^\infty dr^\prime \frac{\sqrt{c(r^\prime)}-1}{r^\prime}.
\end{equation}
Note that we have viewed a vortex as a point defect and ignored its self-energy. When considering the self-energy of a vortex, there exists an additional geometric interaction, which is always smaller than the geometric interaction we considered (See details in SM \cite{Append}). 

\begin{figure}[!htb]
\includegraphics[height=3.2cm, width=8.2cm, angle=0]{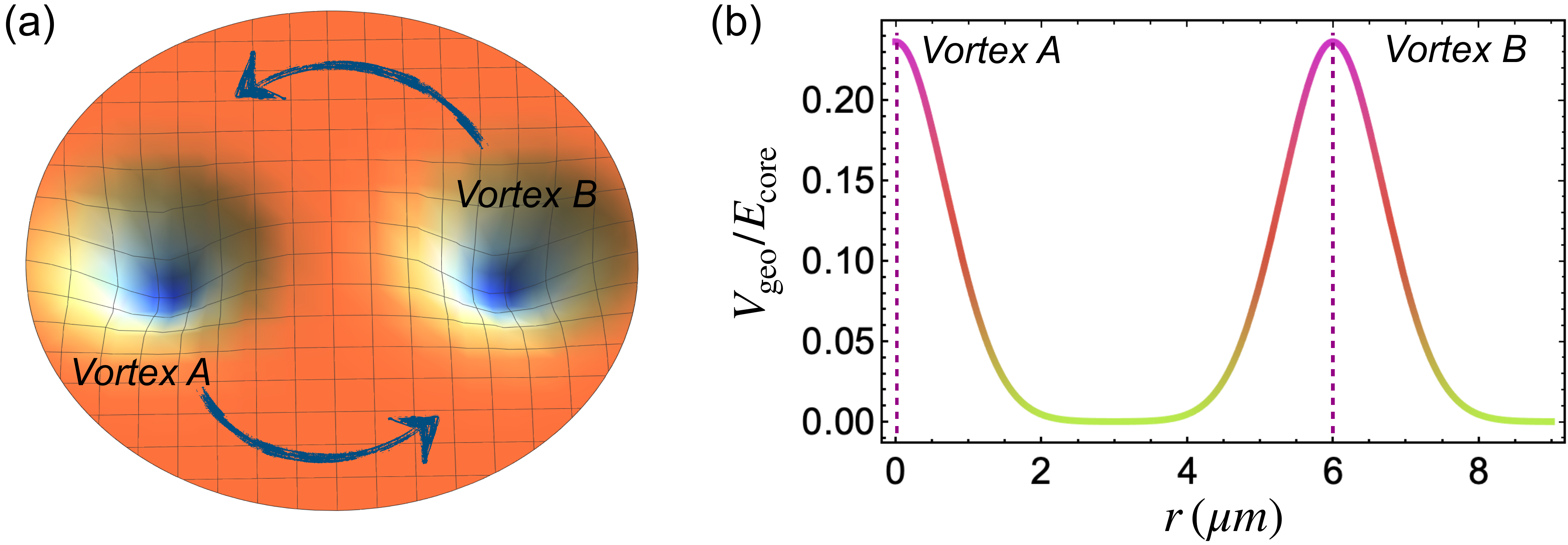}
\caption{(a) Schematic demonstration of quantum braiding by engineering geometric curvature. (b) shows the the geometric potential versus distance with aspect ratio $\alpha=0.8$ for each valley.
\label{fig3}}
\end{figure}

The vortex-geometry interaction provides a unique route to control the position of a vortex. And since a localized Majorana mode is associated with a vortex in a chiral superfluid, one can adiabatically braid Majorana modes by mechanically engineering geometric curvature, as is illustrated in Fig. \ref{fig3} (a). We plot the geometric potential (for vortices) generated by two valleys in Fig. \ref{fig3} (b). It shows that the geometric potential is comparable to the self energy of a vortex. Therefore, the vortex-geometry interaction offers a promising route to perform topological quantum computing in the future.

{\bf Anomalous mass and spin supercurrent in $\rm ^3$He superfluid thin film.---}
We apply geometric induction theory in chiral superfluid $^3$He film. While both $^3$He and $^4$He are superfluids at sufficiently low temperature, the superfluidity in $^3$He is more closely resembles superconductivity than the superfluid $^4$He. Because, unlike $^4$He, $^3$He atoms are fermions that have to be paired to become superfluid. In $^3$He the strong repulsive force exerted by the atomic cores prevents s-wave pairing: instead, the pairs form an orbital p-wave state, with $L$ and $S$ are both equal to $\hbar$. We will consider the $^3$He-A phase where Cooper pairs possess finite angular momentum in the z-direction $L_z$. Near a surface, surface scattering favors the orbital angular momentum $L_z$ perpendicular to the surface \cite{IkegamiTK}. As a result, our geometric induction theory applies.

In $^3$He-$\rm A$ ($\rm A_1$, $\rm A_2$) phase the spin-up and spin-down components have the same chirality, and the corresponding order parameter reads \cite{Vollhardt}
\begin{equation}\label{ophea}
\Psi_A= \frac{1}{\sqrt{2}}\left(\mathbf{\hat e_x}+i\,\mathbf{\hat e_y}\right)
\left(\sqrt{\rho_\uparrow}e^{i\theta_\uparrow}|\uparrow\rangle+\sqrt{\rho_\downarrow}e^{i\theta_\downarrow}|\downarrow\rangle\right)
\end{equation}
where $\rho_{\uparrow/\downarrow}$ and $\theta_{\uparrow/\downarrow}$ are the superfluid density and phase of the spin-up/down component, respectively. Depending on the relative magnitude of $\rho_\uparrow$ and $\rho_\downarrow$, this order parameter can describe $^3$He-A phase ($\rho_\uparrow=\rho_\downarrow$),  $\rm A_1$ phase (either $\rho_\uparrow$ or $\rho_\downarrow$ vanishes),  or $\rm A_2$ phase ($\rho_\uparrow\neq\rho_\downarrow$).
Assuming constant superfluid density, we can obtain the Ginzburg-Landau (GL) Lagrangian density for $^3$He superfluid thin film embedded on a curved surface
\begin{eqnarray}\label{laghea}
\mathcal L_A&=&\frac{\gamma_\uparrow}{2}\left(\partial_\mu \theta_\uparrow+{\omega_\mu}+\mathcal A^{ac}_\mu\right)^2+\frac{\gamma_\downarrow}{2}\left(\partial_\mu \theta_\downarrow+{\omega_\mu}-\mathcal A^{ac}_\mu\right)^2\nonumber\\
&&+ \text{interacting terms + potential terms...}
\end{eqnarray}
where $\gamma_{\uparrow/\downarrow}=\frac{\rho_{\uparrow/\downarrow}}{m}$ denotes the stiffness for spin-up/down component; $\left( \mathcal A^{ac}_0, {\mathcal A^{ac}_k}\right)=\left(\mu_i B_i, \varepsilon_{ijk}E^{i}\mu^j\right)$ is the Aharonov-Casher (AC) gauge field arising due to a magnetic moment $\boldsymbol \mu$ moving in an electromagnetic field $\left(\mathbf E,\mathbf B\right)$ \cite{AharonovC,Shen}.

One can obtain the current density of the spin-up and spin-down components from the Lagrangian density $\mathcal L_A$
$j_\mu^{\uparrow/\downarrow}=\gamma_{\uparrow/\downarrow}\left[\partial_\mu\theta_{\uparrow/\downarrow}+\omega_\mu\pm{\mathcal A^{ac}_\mu}\right]$.
Defining a total mass current $j_\mu^{\text{m}}=j_\mu^\uparrow+j_\mu^\downarrow$ and a total spin current $j_\mu^{\text{s}}=j_\mu^\uparrow-j_\mu^\downarrow$ yields the matrix formula:
\begin{eqnarray}\label{eqmatrixform1}
\left(
\begin{array}{cc}
j_\mu^{\text{m}}\\  j_\mu^{\text{s}}
\end{array}
\right)
=\left(
\begin{array}{cc}
\gamma^{\text{m}} &\gamma^{\text{s}}\\
\gamma^{\text{s}}&\gamma^{\text{m}}
\end{array}
\right)\cdot
\left(
\begin{array}{cc}
\omega_\mu\\{\mathcal A^{ac}_\mu}
\end{array}
\right),
\end{eqnarray}
where $\gamma^{\text{m/s}}\equiv \gamma_\uparrow\pm\gamma_\downarrow$,
and the phase gradient term is absorbed into the $\omega_\mu$ and $\mathcal A_\mu^{ac}$ by a gauge transformation.
One can immediately make several useful predictions from Eq.\eqref{eqmatrixform1}. In $\rm ^3He$-A phase $\gamma^{\text{s}}=0$ indicates that Gaussian curvature drives a mass current whereas the AC gauge field drives a spin current.  In $\rm ^3He$-$\rm A_1$ or $\rm A_2$ phase, however, $\gamma^{\text{s}}$ is finite so that  either Gaussian curvature or an AC gauge field can drive both mass current and spin current, simultaneously.
Generally, symmetry should allow a spin-spin interaction term \cite{Leggett1968} such as $j_\mu^{\uparrow}\, j^{\mu\downarrow}$. As is discussed in detail in SM \cite{Append}, the spin-spin interaction effectively shifts the strength of mass or spin stiffness.

{\bf Electromagnetic signature.}---
We obtain the electromagnetic signature of chiral superfluids induced by geometric gauge fields, and for definiteness we take $\rm ^3He$-A phase as an example. Minimization of GL action with respect to the four-vector potential $A_\mu=\left(\phi, \mathbf A\right)$ leads to the effective electric charge and electric current density
\cite{Append}:
\begin{eqnarray}\label{emsigma}
\sigma_{c}=-\gamma^{\rm s}\,\mu\, \mathcal B(\mathbf r),~~~
\bold J_{c}=\gamma^{\rm s}\,\boldsymbol \mu \times\boldsymbol{\mathcal E}(\mathbf r)
\end{eqnarray}
where $\mathcal B(\mathbf r)$ and $\boldsymbol{\mathcal E}(\mathbf r)$ are the magnetic-like and electric-like geometric field strength in Eq.\eqref{GInduction}; $\boldsymbol\mu=\mu\bold{\hat e_3}$ is the magnetic moment perpendicular to the surface.
The definition of the geometric field strength leads to the the Maxwell-like equation $\boldsymbol{\nabla}\times\boldsymbol{\mathcal E}=\partial_t \mathcal B$, which further guarantees the current conservation $\partial_t \sigma_c+\boldsymbol{\nabla}\cdot \mathbf J_c=0$.
One observes that effective charge density and electric current density can emerge when there is a stiffness difference between the spin-up component and the spin-down component. Similar reasoning enables us to obtain the effective electric charge density and current density for several other chiral phases of $^3$He \cite{Append}. We assume a superfluid density $\rho\approx 10^{22}/{\rm m^2}$ and a Gaussian curvature $\mathcal B\approx 1/(100{\rm \mu m})^2$. The effective charge density can induce an electric field $E\approx 10^{-3}{\rm V/m}$.

{\bf Geometric induction in a flexible superfluid thin film.}---
We consider the geometric induction theory of a chiral superfluid embedded on a flexible surface. The flexibility of the surface provides additional degrees of freedom to minimize the total GL action:
\begin{eqnarray}
S_{\text{tot}}
=\int dt d^2 r\sqrt{g}\,\,&&\left\{\frac{\gamma}{2}\right.\left(\partial_\mu \theta+\ell\,\omega_\mu\right)^2\nonumber\\
&&\left.+\left[\frac{\kappa_0}{2}\left(\partial_t h\right)^2-\frac{\kappa_r}{2} \left(\nabla^2 h\right)^2\right] \right\}
\end{eqnarray}
where the first term and the second term represent the the Lagrangian of chiral superfluid and geometry, respectively. To describe a flexible surface, we use height $h(x,y,t)$ - the deviation of a curved surface from a plane - to parametrize a 2D surface. The geometric stiffness $\kappa_0$ and $\kappa_r$ measure the softness of the surface \cite{Nelson}. The geometric connection $\omega_\mu=\frac{1}{2}\varepsilon^{0\beta\gamma}\partial_\gamma\left(\partial_\beta h\partial_\mu h\right)$ embodies the essential interaction between a chiral superfluid and geometry.
Minimizing the GL action with respect to the  $h$, one obtains the equation of motion for geometry to linear order in height and supercurrent density $j_\mu$:
\begin{eqnarray}
\kappa_0 \partial^2_t h-\kappa_r\nabla^4 h = \ell \,\left(\partial_\mu\partial_\beta h\right)\varepsilon^{0\beta\gamma}\partial_\gamma j^\mu.
\end{eqnarray}

\begin{figure}[!htb]
\includegraphics[height=4.2cm, width=6.2cm, angle=0]{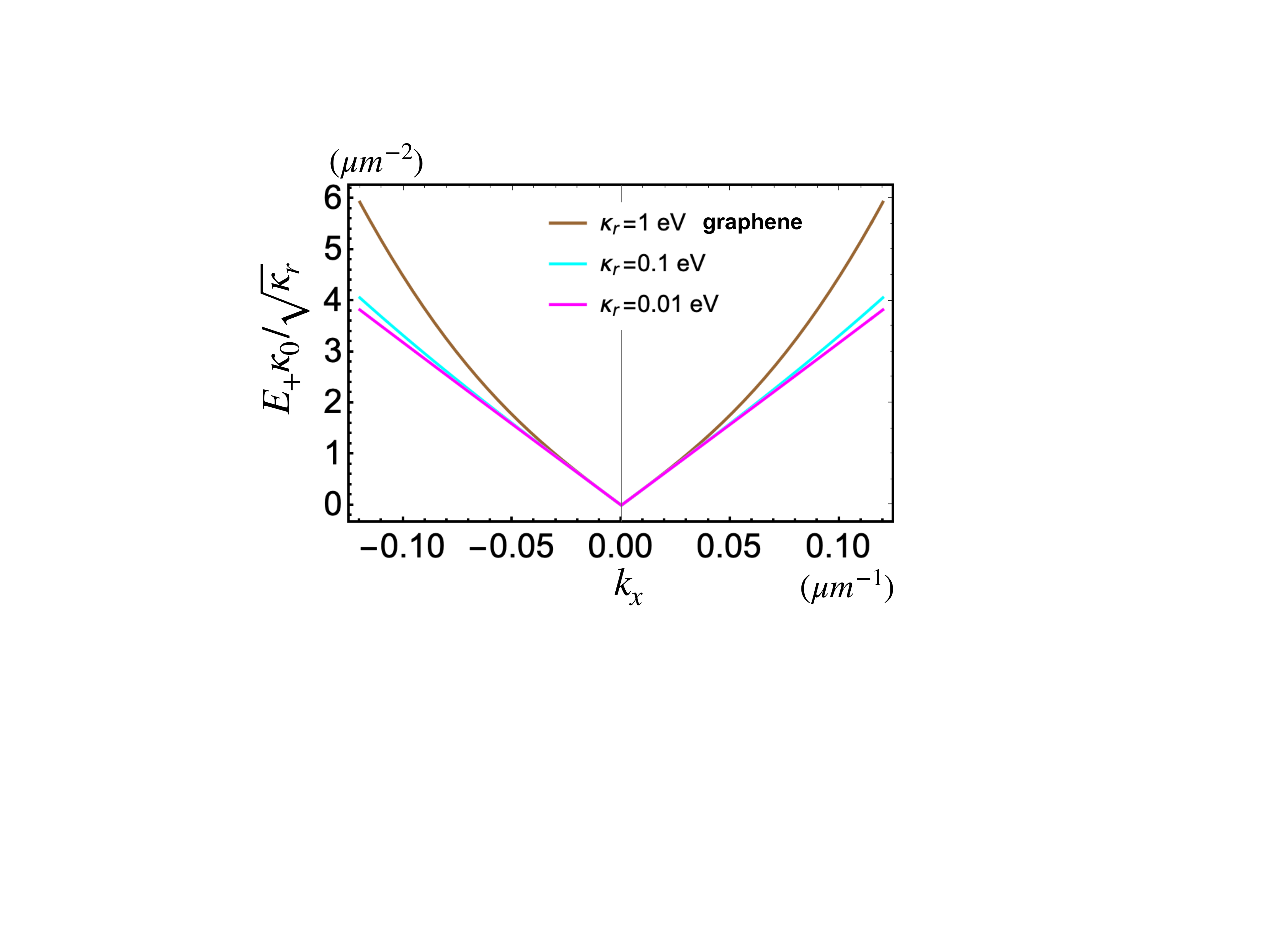}
\caption{The energy-momentum dispersion of $h$ is shown for three stiffness $\kappa_r$. For comparison, a suspended graphene has a stiffness $k_r=1{\rm eV}$. Numerically, we have assumed a reasonable superfluid density $\rho\approx 10^{22}/{\rm m^2}$ and superfluid current gradient $\Gamma=10^{-7}{\rm J/m^2}$. We set $\ell=1$ and $k_y=0$ in the plot.
\label{fig4}}
\end{figure}
The dynamic of geometry is qualitatively modified due to the presence of chiral superfluid. To quantify the influence of chiral superfluidity on geometry, we study the energy-momentum dispersion of $h$ (usually called flexural modes) by assuming a supercurrent in the x-direction with a gradient $\Gamma\equiv\partial_y j^x$ in the y-direction. We obtain the modified dispersion relation due to the backaction of the chiral superfluid:
\begin{eqnarray}
E_{\pm}=\pm \sqrt{\frac{\kappa_r}{\kappa_0}\left(k_x^2+k_y^2\right)^2+\frac{\ell \,\Gamma}{\kappa_0} k_x^2}.
\end{eqnarray}
In Fig. 4 we see that the energy-momentum dispersion in the x-direction becomes Dirac type at small momentum, i.e., $k_x\ll k_c\equiv  \sqrt{|{\ell\Gamma}/{\kappa_r}|}$, with the critical speed  $v_{c}= \sqrt{\ell \,\Gamma/\kappa_0}$. Given $\ell=1$, $\Gamma=10^{-7}{\rm J/m^2}$, and the geometric stiffness $\kappa_0\approx 7.6\times 10^{-8}{\rm g/cm^2}$ (values taken from graphene\cite{flxuralGraphene}), we can estimate the emergent critical speed $v_c\approx 0.36 {\rm m/s}$.

{\bf Summary.---}
We have studied the intriguing interplay between chiral superfluidity and geometry. Due to the chiral order parameter, a geometric gauge field emerges and induces anomalous dynamics and interactions in chiral superfluids. Based on the anomalous interaction between vortices and geometry, we proposed a mechanical approach to control the positions of vortices, which creates a new route for quantum braiding. We further show that both mass supercurrent and spin supercurrent can be driven by a Gaussian curvature. And we also obtained the geometry-induced electromagnetic signatures. Finally, we study the backaction of chiral superfluidity on geometry. We find that the dispersion of geometry shifts from quadratic to linear due to the presence of chiral superfluidity.

Several proposed effects illustrate the opportunities of controlling quantum states with strain, e.g., pseudo-electromagnetic fields in 3D topological semimetals \cite{Rllan}, uniaxial pressure control of competing orders in a high-temperature superconductor \cite{HKim}, strain and ferroelectric soft-mode induced superconductivity in strontium titanate \cite{KDunnett}. While it is known that strain can affect the superconducting state, this work highlights the opportunities to induce and modify spin and mass currents in superfluids.

\textit{Acknowledgement}: 
We are grateful for useful discussions and previous collaborations with T. H. Hansson and F. Wilczek.  Q.-D. Jiang was sponsored by Pujiang Talent Program 21PJ1405400 and TDLI starting up grant. AB was supported by  the European Research Council ERC HERO-810451 grant, University of Connecticut, and the Swedish Research Council (VR).%


\clearpage
\begin{widetext}
\appendix

\begin{center}
\textbf{Supplemental Materials}
\end{center}

\begin{center}
\textbf{S-I. Mathematical description of geometric connection.}
\end{center}

In this section, we deal with tangent-plane order on a curved surface by reviewing relevant concepts in differential geometry, mostly to establish notations. In standard literature, what we called geometric connection is actually called spin connection although it has nothing to do with real spin. To avoid confusion, we only call it geometric connection in the main text. However, we recover its standard name ``spin connection" in the supplementary for the sake of mathematical consistence. 

\textbf{Differential geometry of a two-dimensional surface.}---
A two-dimensional surface embedded in three-dimensional Euclidean space can be parametrized by a three-dimensional vector $\bold R(\bold r)=(R_1(\bold r), \,R_2(\bold r),\,R_3(\bold r))$, as a function of a two-dimensional parameter $\bold r=(x^1,x^2)$. Covariant tangent-plane vectors are defined as 
\begin{eqnarray}
\bold t_\alpha=\partial_\alpha \bold R, ~ \alpha=1,2,
\end{eqnarray}
where $\partial_\alpha=\partial/\partial x^\alpha$ with $\alpha,\,\beta...$ to denote components of vectors and tensors written in the local coordinates.
The metric tensor is 
$
g_{\alpha}=\bold t_\alpha\cdot \bold t_\beta
$, and the square root of the determinant of $g_{\alpha\beta}$, $\sqrt{g}=\sqrt{{\rm det}\, g_{\alpha\beta}}$, is useful for constructing invariant area. 
$g^{\alpha\beta}$, the inverse of $g_{\alpha\beta}$, is defined as
$g^{\alpha\beta}g_{\beta\gamma}=\delta_\gamma^\alpha$. One can define contravariant tangent-plane vectors $\bold t^\alpha=g^{\alpha\beta}\bold t_\beta$ satisfying $\bold t^\alpha\cdot\bold t_\beta=\delta^\alpha_\beta$. Any vector $\bold V$ in the tangent plane can be expressed as $\bold V=V^\alpha \bold t_\alpha=V_\alpha \bold t^\alpha$, where $V_\alpha=\bold V\cdot t_\alpha$ and $V^\alpha=\bold V\cdot \bold t^\alpha=g^{\alpha\beta}V_\beta$ are the covariant and contravariant components of $\bold V$. 

A unit $\bold{\hat n}$ normal to the surface can be constructed from $\bold t_1$ and $\bold t_2$, namely
$\bold{\hat n}=\frac{\bold t_1\times\bold t_2}{|\bold t_1\times\bold t_2|}$. And the curvature tensor can be obtained via the formula
\begin{eqnarray}
K_{\alpha\beta}=\bold{\hat n}\cdot \partial_\alpha\partial_\beta \bold R.
\end{eqnarray}
which is very useful for calculating the mean (extrinsic) curvature $H=\frac{1}{2}K_\alpha^\alpha$ 
and the Gaussian (intrinsic) curvature $\mathcal B={\rm det} \, K_\beta^\alpha={\rm det}\,(g^{\alpha\gamma}K_{\gamma\beta})={\rm det}\,K_{\gamma\beta}\,{\rm det} g^{\alpha\gamma}$.

To be specific, let us calculate the Gaussian curvature of a curved surface parametrized in the Monge representation, i.e., $\bold r=(x^1,x^2)$ and $\bold R(\bold r)=(\bold r,\,h(\bold r))$. In the Monge representation, one can obtain the metric tensor $g_{\alpha\beta}$ and the inverse metric tensor $g^{\alpha\beta}$:
\begin{eqnarray}
g_{\alpha\beta}=\partial_\alpha \bold R\cdot \partial_\beta\bold R=\left(
\begin{array}{cc}
1+(\partial_1 h)^2 &\partial_1 h\partial_2 h\\
\partial_1 h\partial_2 h& 1+(\partial_2 h)^2
\end{array}\right),
\end{eqnarray}
\begin{eqnarray}
g^{\alpha\beta}=g_{\alpha\beta}^{-1}=\frac{1}{1+(\nabla h)^2}\left(
\begin{array}{cc}
1+(\partial_2 h)^2 &-\partial_1 h\partial_2 h\\
-\partial_1 h\partial_2 h& 1+(\partial_1 h)^2
\end{array}\right),
\end{eqnarray}
where $(\nabla h)^2=(\partial_1 h)^2+ (\partial_2 h)^2$. Based on the matrix form of metric tensor, one can conveniently write $g_{\alpha\beta}=\delta_{\alpha\beta}+(\partial_\alpha h)(\partial_\beta h)$ and $g^{\alpha\beta}=\delta_{\alpha\beta}-(\partial_\alpha h) (\partial_\beta h)$ to the second order approximation.
The curvature tensor $K_{\alpha\beta}$ is
\begin{eqnarray}
K_{\alpha\beta}=\bold n \cdot \partial_\alpha\partial_\beta \bold R=\frac{1}{\sqrt{1+(\nabla h)^2}}
\left(\begin{array}{cc}
\partial_1^2 h & \partial_1\partial_2 h\\
\partial_1\partial_2 h & \partial_2^2 h
\end{array}
\right).
\end{eqnarray}
Consequently, the Gaussian curvature and mean curvature can be derived via the following formulas:
\begin{eqnarray}
\mathcal B={\rm det} K_{\alpha\beta} \,{\rm det}\, g^{\beta \gamma}=\frac{|K_{\alpha\beta}|}{|g_{\beta\gamma}|}=\frac{1}{[1+(\nabla h)^2]^2}\left[\partial_1^2h\partial_2^2h-(\partial_1\partial_2 h)^2\right].
\end{eqnarray}
\begin{eqnarray}
H=K_{\alpha\beta} g^{\beta\alpha}=\frac{1}{[1+(\nabla h)^2]^{\frac{3}{2}}}\left\{[1+(\partial_2 h)^2]\partial_1^2 h+[1+(\partial_1 h)^2]\partial_2^2 h-2(\partial_1\partial_2h)(\partial_1h\partial_2 h)\right\}.
\end{eqnarray}

For later convenience, we need to define the antisymmetric tensor $\varepsilon_{\alpha\beta}$ via
\begin{eqnarray}
\varepsilon_{\alpha\beta}=\bold{\hat n} \cdot (\bold t_\alpha\times \bold t_\beta)=\sqrt{g}\epsilon_{\alpha\beta},
\end{eqnarray}
where $g=det\, g_{\alpha\beta}$ and $\epsilon_{\alpha\beta}$ is the antisymmetric tensor with $\epsilon_{12}=-\epsilon_{21}=1$. The contravariant tensor is
\begin{eqnarray}
\varepsilon^{\alpha\beta}=\bold{\hat n}\cdot (\bold t^\alpha\times \bold t^\beta)=\epsilon_{\alpha\beta}/\sqrt{g},
\end{eqnarray}
and satisfies $\varepsilon^{\alpha\beta}\varepsilon_{\beta\gamma}=-\delta^\alpha_\gamma$. Finally, the mixed tensor
$\varepsilon^\alpha_\beta=g^{\alpha\gamma}\varepsilon_{\gamma\beta}$
rotates a vector by $\pi/2$, because $V_\alpha\varepsilon^\alpha_\beta V^\beta=\varepsilon_{\alpha\beta}V^\alpha V^\beta=0$ and $\varepsilon^\alpha_\beta V^\beta \varepsilon_\alpha^\gamma V_{\gamma}=V^\alpha V_\alpha$.

\textbf{Description of the tangent-plane order.---}
We focus on the situation where the tangent-plane order has a fixed magnitude. Therefore, it is useful to introduce a set of orthonormal tangent-plane basis vectors $\bold e_1$ and $\bold e_2$ satisfying
\begin{eqnarray}
\bold{\hat e_a}\cdot \bold{\hat e_b}=\delta_{ab}, ~ \bold{\hat n}\cdot \bold{\hat e_a}=0; ~ (a=1,2)
\end{eqnarray}
A tangent vector $\bold V$ can be expressed in the basis $\{\bold{\hat e_1}, \,\bold{\hat e_2}\}$ as well as that defined by the covariant or contravariant vectors: $\bold V=V_a \bold{\hat e_a}$ where $V_a=\bold{\hat e_a}\cdot \bold V$. Notice that the covariant derivatives are the derivatives projected into the tangent plane. Components of the covariant derivative of a vector $\bold V$ relative to the orthonormal basis are 
\begin{eqnarray}
D_\alpha V_a\equiv \bold{\hat e_a}\cdot (\partial_\alpha \bold V)=\partial_\alpha V_a+\bold{\hat e_a}\cdot \partial_\alpha \bold{\hat e_b} \,V_b=\partial_\alpha V_a+\epsilon_{ab} \omega_{\alpha}\, V_b,
\end{eqnarray}
where $\omega_\alpha=\bold{\hat e_1}\cdot \partial_\alpha \bold{\hat e_2}$ is the spin-connection whose curl is the Gaussian curvature, i.e.,$
\varepsilon^{\alpha\beta}\partial_\alpha \omega_\beta=\mathcal B$. 
Note that it is $\varepsilon$ instead of $\epsilon$ in the expression of the above formula. We need to use $\varepsilon$ because we want to calculate the curl in the normal direction of the surface, not the curl in the z-direction.  

\textbf{Spin connection and affine connection.---} It is instructive to see how to obtain $\boldsymbol{\omega}$ \cite{MBowick2009a}. To describe a tangent-plane order, one need to define a local tangent-plane orthonormal coordinate frame (a tetrad, or a vielbein) satisfying
\begin{eqnarray}
\bold{\hat e_a}=e_a^\alpha \bold t_\alpha;~ \bold e_b=e_b^\beta\bold t_\beta,
\end{eqnarray}
where $\bold t_\alpha$ and $\bold t_\beta$ are the local tangent-plane vectors defined before. (Note that $\bold t_\alpha$ and $\bold t_\beta$ are generally not unit vectors.) We require $\bold e_a$ and $\bold e_b$ to be orthonormal, which means $\bold e_a\cdot\bold e_b=\delta_{ab}$. This equally indicates 
\begin{eqnarray}\label{etaabac}
\delta_{ab}&=&\bold{\hat e_a}\cdot \bold{\hat e_b}\nonumber\\
&=&(e_a^\alpha \bold t_\alpha)\cdot(e_b^\beta \bold t_\beta)=e_a^\alpha\,e_b^\beta \,g_{\alpha\beta}.
\end{eqnarray}
The spin connection is then obtained from
\begin{eqnarray}\label{spinconnectionac}
\bold{\hat e_a}\cdot \partial_\alpha \bold{\hat e_b}&=&(e_a^\beta \bold t_\beta)\cdot\partial_\alpha\,(e_b^\gamma \bold t_\gamma)\nonumber\\
&=&(e_a^\beta \partial_\alpha e_b^\gamma )\,g_{\beta\gamma}+e_a^\beta e_b^\gamma\, (\partial_\alpha\,\bold t_\gamma\cdot\bold t_\beta)\nonumber\\
&=&(e_a^\beta \partial_\alpha e_b^\gamma )\,g_{\beta\gamma}+e_a^\beta e_b^\gamma\,\Gamma_{\beta \gamma \alpha},
\end{eqnarray}
where $\Gamma_{\beta \alpha\gamma}$ is the Christoffel symbols (affine connection) of the first kind. The Christoffel symbols $\Gamma_{\nu \lambda}^{\mu}$ are used to specify the parallel transport in a tetrad-free language: $V^{\beta}\rightarrow V^\beta-\Gamma^{\beta}_{\alpha\gamma}dx^{\gamma} V^{\alpha}$ \cite{FrancescoMathieua}. In fact, $\Gamma_{\beta\alpha\gamma}$ can be obtained directly from the metric alone, via 
\begin{eqnarray}
\Gamma_{\beta\alpha\gamma}=\frac{1}{2}\left(\frac{\partial g_{\beta\alpha}}{\partial x^\gamma}+\frac{\partial g_{\beta\gamma}}{\partial x^\alpha}-\frac{\partial g_{\alpha\gamma}}{\partial x^\beta}\right).
\end{eqnarray}
One should notice that the above procedures still cannot uniquely determine $\boldsymbol{\omega}$. One can decide a particular form of $\boldsymbol{\omega}$ only after choosing a particular set of $\{e_a^\alpha,\,e_b^\beta\}$ satisfying Eq.\eqref{etaabac}. (The reason is quite physically intuitive. Affine connection is a connection for close tangent planes; while spin connection here is a connection for vectors in tangent planes. Even tangent plane is fixed, one still has the freedom to choose the local coordinate frame.) However, the Gaussian curvature is a gauge-invariant quantity that doesn't depend on what particular frame you choose. With the above procedures, we can, for sure, get one form of spin connection $\boldsymbol{\omega}$ in the Monge representation.

Let's work out the spin connection in the Monge representation. Due to the expression of metric $g_{ij}$, we can derive the Christoffel symbols
\begin{eqnarray}\label{christoffelac}
\Gamma_{\beta\alpha\gamma}&=&\frac{1}{2}\left(\partial_\gamma g_{\beta\alpha}+\partial_\alpha g_{\beta\gamma}-\partial_\beta g_{\alpha\gamma}\right)\nonumber\\
&=&\frac{1}{2}\left\{\partial_\gamma[(\partial_\beta h)(\partial_\alpha h)]+\partial_\alpha[(\partial_\beta h)(\partial_\gamma h)]-\partial_\beta[(\partial_\alpha h)(\partial_\gamma h)]\right\}\nonumber\\
&=&(\partial_\beta h)\,(\partial_\alpha\partial_\gamma h)
\end{eqnarray}
Next, we choose vielbein that satisfies equation \eqref{etaabac}. If we write  \eqref{etaabac} in an explicit manner
\begin{eqnarray}
\left(\begin{array}{cc}
e_1^1& e_1^2\\
e_2^1 & e_2^2
\end{array}
\right)
\left(\begin{array}{cc}
1+(\partial_1 h)^2&(\partial_1 h)(\partial_2 h)\\
(\partial_1 h)(\partial_2 h) & (1+\partial_2 h)^2
\end{array}
\right)
\left(\begin{array}{cc}
e_1^1& e_2^1\\
e_1^2 & e_2^2
\end{array}
\right)=
\left(\begin{array}{cc}
1& 0\\
0 & 1
\end{array}
\right)
\end{eqnarray}
Notice that there are in total 4 unknowns, $e_{1}^1$, $e_{1}^2$, $e_{2}^1$, $e_{2}^2$, while only three equations. (It looks like that we have 4 equations, but two of them are identical.)
\begin{subequations}
\begin{eqnarray}
e_1^1 e_1^1 [1+(\partial_1 h)^2]+e_1^1 e_1^2 [(\partial_1 h)(\partial_2 h)]+e_1^1 e_1^2 [(\partial_1 h)(\partial_2 h)]+e_1^2e_1^2 [1+(\partial_2 h)^2]=1\\
e_1^1 e_2^1 [1+(\partial_1 h)^2]+e_1^2 e_2^1 [(\partial_1 h)(\partial_2 h)]+e_2^2 e_1^1 [(\partial_1 h)(\partial_2 h)]+e_1^2e_2^2 [1+(\partial_2 h)^2]=0\\
e_2^1 e_2^1 [1+(\partial_1 h)^2]+e_2^1 e_2^2 [(\partial_1 h)(\partial_2 h)]+e_2^2 e_2^1 [(\partial_1 h)(\partial_2 h)]+e_2^2e_2^2 [1+(\partial_2 h)^2]=1
\end{eqnarray}
\end{subequations}
If we choose the gauge $e_2^1=e_1^2$, then solving the above equations yields
\begin{eqnarray}\label{veilbeinac}
e_1^1=1-\frac{1}{2} (\partial_1 h)^2; ~e_2^2=1-\frac{1}{2} (\partial_2 h)^2; ~e_1^2=e_2^1=-\frac{1}{2}(\partial_1 h)(\partial_2 h).
\end{eqnarray}
Substitute Eqs.\eqref{christoffelac} and \eqref{veilbeinac} into Eq.\eqref{spinconnectionac}, we can obtain the spin connection 
\begin{eqnarray}
\mathcal W_\alpha&=&\bold e_1\cdot \partial_\alpha \bold e_2\nonumber\\
&=&e_1^\beta\partial_\alpha e_2^\gamma g_{\beta\gamma}+e_1^\beta e_2^\gamma \Gamma_{\beta\gamma\alpha}\nonumber\\
&=& -\frac{1}{2}[(\partial_1 \partial_\alpha h)\partial_2 h+\partial_1 h(\partial_2 \partial_\alpha h)]+\partial_1 h\,(\partial_2\partial_\alpha h)\nonumber\\
&=&\frac{1}{2}\epsilon^{\beta \gamma} [(\partial_\beta h)(\partial_\gamma\partial_\alpha h)]=\frac{1}{2}\epsilon^{\beta\gamma} \partial_\gamma [(\partial_\beta h)\, (\partial_\alpha h)],
\end{eqnarray}
where we have neglected terms of order higher than $(\nabla h)^2$. Note that this is consistent with the result in Ref. \cite{SeungNelson}.

\textbf{Spin connection in chiral basis.---} 
We will also find it useful to use a circular basis defined by the vectors 
\begin{eqnarray}
\boldsymbol \epsilon_{\pm}=\frac{1}{\sqrt{2}}(\bold{\hat e_1}\pm i\bold{\hat e_2})=\boldsymbol \epsilon_{\mp}^*
\end{eqnarray}
satisfying $\boldsymbol \epsilon_a\cdot \boldsymbol \epsilon_b^*=\delta_{ab}$ with $a,\, b=\pm$. In this basis, $\bold V=\tilde{V}_a \boldsymbol\epsilon_a^*$, and the covariant derivative,
\begin{eqnarray}
D_\alpha \tilde{V}_{\pm}\equiv&&\boldsymbol \epsilon_{\pm}\cdot \partial_\alpha \bold V=\partial_\alpha \tilde{V}_{\pm}+\boldsymbol \epsilon_{\pm}\cdot \partial_\alpha \boldsymbol \epsilon_{a}^* \,V_a\nonumber\\
=&&\partial_\alpha \tilde{V}_{\pm}\mp i\omega_\alpha \tilde{V}_{\pm}
=(\partial_\alpha \mp i\omega_\alpha)\tilde{V}_{\pm}
\end{eqnarray}
has a particular simple form \cite{Park1996}.

\textbf{The physical meaning of spin connection and affine connection.---} 
We have shown that spin connection can be expressed in terms of affine connection. Let us again see the difference between affine connection (Christoffel symbols) and spin connection in terms of their physical meanings. 
One can write a vector either in terms of tangent-plane coordinates ($\bold t_\alpha, \bold t_\beta$) or in terms of local coordinates ($\bold{\hat e_a}, \bold{\hat e_b}$):

i) If we choose to express a vector in terms of tangent-plane coordinates (non-orthogonal and non-unit), we can get
\begin{equation}
\begin{aligned}
\partial_\alpha \bold V=&\partial_\alpha [V^\beta \bold t_\beta]\\
=&[\partial_\alpha V^\beta]\bold t_\beta+V^\beta\partial_\alpha \bold t_\beta\\
=&[\partial_\alpha V^\beta]\bold t_\beta+V^\beta \Gamma_{\beta\alpha}^\gamma \bold t_\gamma+V^\beta N_{\beta\alpha}\hat n
\end{aligned}
\end{equation}
In the second line, the second term expresses the effect that the basis vectors themselves vary as we move about. One should notice that $\partial_\alpha \bold V$ contains a component along $\hat n$, the normal to the surface. The Christoffel symbols only include the information of basis vector variation in the tangent plane. The definition of covariant derivative on components defined in global coordinate is 
\begin{equation}
\begin{aligned}
D_\alpha V^\beta\equiv &\bold t^\beta \cdot \partial_\alpha \bold V=\bold t^\beta \cdot \partial_\alpha (V^\gamma \bold t_\gamma)\\
=&\partial_\alpha V^\beta+\Gamma_{\gamma\alpha}^\beta V^\gamma,
\end{aligned}
\end{equation}
where the out-of-tangent-plane component is projected out. 

ii) If we choose to express a vector in terms of local orthonormal coordinates, we get
\begin{equation}
\begin{aligned}
\partial_\alpha \bold V=&\partial_\alpha[V^a \bold{\hat e_a}]
=[\partial_\alpha V^a]\hat e_a+V^a\partial_\alpha \bold{\hat e_a}\\
=&[\partial_\alpha V^a]\hat e_a+V^c \omega_{c \alpha}^b\bold{\hat e_b}.
\end{aligned}
\end{equation}
The definition of covariant derivative on components defined in the local frame is 
\begin{equation}
D_\alpha V^a\equiv \bold{\hat e_a} \cdot \partial_\alpha \bold V
=\partial_\alpha V^a+\omega_{c\alpha}^a V^c,
\end{equation}
where $\omega_{c \alpha}^a$ is the spin connection. That is the spin connection connects vectors defined in local orthomormal coordinates $\bold{\hat e_\alpha}$ while affine connection connects vectors defined in terms of $\bold t_\alpha$. 

\begin{center}
\textbf{S-II. Spin connection as a gauge field.}
\end{center}

In the main text, we have shown the emergence of a geometric phase when parallel transporting a complex vector. Here, we give the mathematical derivations of the geometric gauge field arising from transporting a complex tensor.

\textbf{Method 1.}---
One may first understand the appearance of geometric gauge field through a hand-waving argument as follows:
With the assumption of constant superfluid density $\rho$, the chiral order parameter is totally determined by its local phase, i.e., $\Psi(\bold r)=(\psi_{x}\pm i\psi_{y})^\ell=\sqrt{\rho}\langle exp[\pm i\ell \theta(\bold r)] \rangle$, where $\ell \hbar$ is the angular momentum of a Cooper pair. Note that since the local $U(1)$ phase $\theta(\bold r)$ depends on the choice of orthonormal vectors $\bold{\hat e_1}$ and $\bold{\hat e_2}$, so does the order parameter $\Psi(\bold r)$. This means that any spatial derivatives for $\Psi$ must be covariant derivatives, namely, $\partial_\mu\rightarrow D_\mu=\partial_\mu+i\ell \omega_\mu$. Here, $\omega_\mu=\bold{\hat e_1}\cdot \partial_\mu \bold{\hat e_2}$ is the spin connection that originates from parallel transporting the position-dependent orthonomal vectors.  

\textbf{Method 2.}---
The order parameter for a chiral $\ell -$wave superfluid can be generically written as
\begin{eqnarray}
\Psi=\sqrt{\frac{\rho_+}{2^\ell }}e^{i\theta_+} \left(\bold{\hat e_1}+i \bold{\hat e_2}\right)^\ell  +\sqrt{\frac{\rho_-}{2^\ell }}e^{i\theta_-}\left(\bold{\hat e_1}-i\bold{ \hat e_2}\right)^\ell .
\end{eqnarray}
where $\rho_{\pm}$ represents the superconducting-carrier density for $\pm$ pairing parity, respectively. Here $\Psi$ is a rank-$\ell $ tensor with its magnitude $|\Psi|=\Psi \cdot \Psi^*$, where the dot means the inner product of two tensors. 

The action for a $\ell -$wave chiral superfluid is
\begin{equation}\label{spinconnectproof1}
S_{sc}=\int dt \int d^2 r \sqrt{g} \left\{i\hbar \Psi^* \cdot \partial_0 \Psi-\frac{\hbar^2 g^{ij}}{2m}(\partial_i \Psi)^*\cdot(D_j \Psi)+V(|\Psi|)\right\}.
\end{equation}
There is no spin connection term in the expression because the spin connection is associated with the components, not the vector itself. 
In the following, we choose the potential term $V(|\Psi|)$ that only favors the $``+"$ chirality, i.e., $\rho_+\neq 0$ and $\rho_-=0$. Our results can be easily generalized to the $``-"$ chirality case, straightforwardly. 

Let us first examine the time-dependent part, $\Psi^*\cdot \partial_0 \Psi$:
\begin{equation}
\begin{aligned}
\Psi^* \cdot D_0\Psi=&\rho_+ \left(i\partial_0\theta\right)+\frac{\rho_+}{2^\ell }(\bold{\hat e_1}-i\bold{\hat e_2})^\ell \cdot\left[\partial_0 (\bold{\hat e_1}+i\bold{ \hat e_2})^\ell \right]\\
=& \rho_+\left(i\partial_0\theta\right)+\frac{\rho_+}{2^\ell }\ell (\bold{\hat e_1}-i \bold{\hat e_2})^\ell \cdot (\bold{\hat e_1}+i\bold {\hat e_2})^{\ell -1}\partial_0 (\bold{\hat e_1}+i\bold {\hat e_2})\\
=& \rho_+ \left(i\partial_0\theta\right)+\frac{\rho_+}{2}\ell (\bold{\hat e_1} -i\bold{ \hat e_2})\cdot \partial_0 (\bold{\hat e_1}+i\bold{\hat e_2})\\
=&\rho_+i\left(\partial_0\theta+ \ell \omega_0\right)
\end{aligned}
\end{equation}
where $\omega_0=\bold{\hat e_1}\cdot \partial_0 \bold{ \hat e_2}$ is the temporal part of the spin connection. 

We then examine the space-dependent part, $(\partial_i \Psi)^*\cdot (\partial_j \Psi)$:
\begin{equation}
\begin{aligned}
(\partial_i \Psi)^*\cdot (\partial_j \Psi)=&\frac{\rho_+}{2^\ell }\partial_i\left[e^{-i\theta_+} (\bold{\hat e_1}-i\bold{ \hat e_2})^\ell \right]\cdot \partial_j\left[e^{i\theta_+} (\bold{\hat e_1}+i\bold{ \hat e_2})^\ell \right]\\
=&\rho_+\left(\partial_i \theta_+\right)\left(\partial_j \theta_+\right)+\frac{\rho_+}{2^\ell }\left(-i\partial_i\theta_+\right) (\bold{\hat e_1}-i\bold{\hat e_2})^\ell \cdot \partial_j (\bold{\hat e_1}+i\bold{\hat e_2})^\ell \\
&+\frac{\rho_+}{2^\ell }\left(i\partial_j\theta_+\right)  (\bold{\hat e_1}+i\bold{\hat e_2})^\ell  \cdot\partial_i(\bold{\hat e_1}-i\bold{ \hat e_2})^\ell +\frac{\rho_+}{2^\ell } \partial_i (\bold{\hat e_1}-i\bold{ \hat e_2})^\ell  \cdot \partial_j (\bold{\hat e_1}+i\bold{ \hat e_2})^\ell \\
=&\rho_+\left(\partial_i \theta_+\right)\left(\partial_j \theta_+\right)+\rho_+\ell (-i\partial_i\theta_+)\left(i\omega_j\right)+\rho_+\ell (i\partial_j\theta_+)\left(-i\omega_i\right)+\rho_+\ell ^2\left(\omega_i \omega_j\right)\\
=&\rho_+\left(\partial_i \theta_++\ell \omega_i\right)\left(\partial_j \theta_++\ell \omega_j\right)
\end{aligned}
\end{equation}
Here $\omega_i=\bold{\hat e_1}\cdot \partial_i \bold{\hat e_2}$ is the spatial part of the spin connection. A key step to accomplish the derivation is that one needs to verify that $\partial_i(\bold{\hat e_1}+i\bold{ \hat e_2})\cdot\partial_j(\bold{\hat e_1}-i\bold{ \hat e_2})=2\omega_i\omega_j$. To prove this, one could insert a unit tensor between two vectors, i.e., $\partial_i(\bold{\hat e_1}+i\bold{ \hat e_2})\cdot(\bold{\hat e_1}\bold{\hat e_1}+\bold{ \hat e_2}\bold{ \hat e_2})\cdot\partial_j(\bold{\hat e_1}-i\bold{ \hat e_2})=(i\bold{\hat e_1}\cdot \partial_i \bold{ \hat e_2})(-i\bold{\hat e_1}\cdot \partial_j \bold{ \hat e_2})+(\bold{ \hat e_2}\cdot \partial_i\bold{\hat e_1})(\bold{ \hat e_2}\cdot \partial_j \bold{\hat e_1})=2\omega_i\omega_j$.

\begin{center}
\textbf{S-III. Comparison with the topological field theory of (fractional) quantum Hall effect.}
\end{center}

We do two things in this section. First, we review the deep analogy between superfluidity and (fractional) quantum Hall (QH) effect, mainly to refresh the reader's memory and to make this section self-contain. Second, we compare the similarity and differences between chiral superfluidity and the QH effect. QH effect shares many similar properties with superfluidity such as no dissipation and vortex soliton excitations \cite{ZHK1989app}. This observation indicates that there might exist a universal theoretical understanding of the two very different phenomena. Indeed, according to composite-boson theory, the QH effect can be understood as the Bose-Einstein condensation of charged bosons \cite{ZHK1989app}. Each electron is viewed as a composition of a charged boson and an odd number of fundamental flux units $\phi_0=h/e$. This can be accomplished by introducing a statistical gauge field $\alpha_\mu$ which is determined by the particle density $\rho(\bold r)$ of charged bosons:
\begin{equation}
\nabla\times \boldsymbol\alpha(\bold r)=(2k+1)\phi_0 \rho (\bold r). ~(\text{$k$ is an integer}).
\end{equation}
This formula means that a unit charge induces a statistical flux $\oint \boldsymbol \alpha\cdot d\bold l=(2k+1)\phi_0$. As one interchanges two of these boson-flux composites, a Aharonov-Bohm phase factor of $\exp\left\{i\frac{e}{\hbar}\int_0^\pi \boldsymbol \alpha \cdot d\bold l\right\}=-1$ is obtained, which correctly reproduces the fermionic statistic of electrons. 

According to the composite-boson theory, fractional filling (odd denominator, in particular) of QH effect can be understood as follows: Zero longitudinal resistivity indicates that the system forms a BEC condensate of charged bosons. According to the Meissner effect, the total gauge field seen by charged bosons (i.e., EM field $A_\mu$ and statistic gauge field $\alpha_\mu$) must be canceled out in the bulk. Therefore, a unit bosonic charge has to associate with $(2k+1)\phi_0$ statistic flux and an opposite $-(2k+1)\phi_0$ magnetic flux. The filling factor is defined as the ratio between electron number and magnetic flux number, i.e., $\nu=\frac{N_e}{N_{\phi_{B}}}=\frac{1}{2k+1}$. 

Beyond the fractional filling factor, the superfluid analogy also provides a useful picture to understand other behaviors in the QH effect. For example, the stability of the Hall plateau may be due to the pinning of statistical vortices induced by adding additional charges into the system \cite{stone1990app}. 

Having reviewed the background, we are ready to compare chiral superfluidity with QH effect. Chiral superfluidity has additional similarities with QH effect because they both couple to geometry. Some of the similarities has been noticed in the literature \cite{StoneRoy2004,MorozHoyos}. To make the comparison concrete, let us start from the effective action that describes the long-range physics of QH liquids on a curved surface \cite{WenZee1992app}:
\begin{equation}\label{EFTQH}
\mathcal S_{\text{QH}}=\int dtd^2r\sqrt{g} \left[\frac{1}{4\pi}\left(\varepsilon^{\mu\nu\rho} \alpha_\mu \partial_\nu a_\rho+2 A_\mu \varepsilon^{\mu\nu\rho}\partial_\nu \alpha_\rho +2s\,\omega_\mu\varepsilon^{\mu\nu\rho} \partial_\nu \alpha_\rho\right)+\alpha_\mu j^\mu\right]
\end{equation}
The action contains several pieces: The first term is the well-known Chern-Simons interaction, describing the topological dynamics of the Hall fluid. The second and third terms describe the coupling of the dynamics to the electromagnetic field and the curvature of the space, respectively. Finally, the last term accounts for the coupling between the gauge potential $a_\mu$ and $j_\mu$, the current of the quasiparticles (or vortices) in QH fluid. The parameter $s$ in the third term is known as the Wen-Zee shift describing the coupling between QH liquid and geometry. Physically, this geometric coupling comes about because an electron in a QH liquid carries a ``orbital spin" due to its cyclotron motion, and acquires an additional Berry phase when moving on a curved surface.

In a chiral superfluid, such as $\rm ^3$He, while the carriers are charge-neutral, they could have finite magnetic dipoles. And an Aharonov-Casher gauge field must be taken into account, as was shown in the main text. {Adding up the Aharonov-Casher term, the effective action for spin-up and spin-down components reads (see the part above Eq.(10) in the main text)
\begin{eqnarray}
\mathcal S^{\uparrow/\downarrow}_{\text{eff}}=\int dtd^2r\sqrt{g}\left[ -\frac{\left(f_{\mu\nu}^{\uparrow/\downarrow}\right)^2}{4\gamma_{\uparrow/\downarrow}} +a^{\uparrow/\downarrow}_\lambda \varepsilon^{\lambda\nu\mu}\partial_\nu \left(\partial_\mu \theta^{\uparrow/\downarrow}_{\text{vortex}}\pm\mathcal A_\mu^{ac}+\ell\,\omega_\mu\right)\right]
\end{eqnarray}
where $f_{\mu\nu}^{\uparrow/\downarrow}=\partial_\mu a^{\uparrow/\downarrow}_\nu -\partial_\nu a^{\uparrow/\downarrow}_\mu$ is the strength tensor of the $a_\mu^{\uparrow/\downarrow}$ field. Note that we have assumed that the spin-up component and the spin-down component have the same chirality (e.g. $\rm ^3$He-A phase). We can derive the effective field theory in terms of mass current and spin current:
\begin{eqnarray}
\begin{aligned}
\mathcal S_{\text{eff}}=&\mathcal S_{\text{eff}}^\uparrow+\mathcal S_{\text{eff}}^\downarrow\\
=&\int dtd^2r\sqrt{g}\left[ -\frac{{\gamma^m}}{2\left({\gamma^m}^2-{\gamma^s}^2\right)} \left(f_{\mu\nu}^m\right)^2
+a_\lambda^m\varepsilon^{\lambda\nu\mu} \partial_\nu\left(\partial_\mu \theta_{\text{vortex}}^m+\ell\,\omega_\mu\right)\right.\\
&-\frac{{\gamma^m}}{2\left({\gamma^m}^2-{\gamma^s}^2\right)} \left(f_{\mu\nu}^s\right)^2+a_\lambda^s\varepsilon^{\lambda\nu\mu} \partial_\nu\left(\partial_\mu \theta_{\text{vortex}}^s+\mathcal A_\mu^{ac}\right)+\left.\frac{{\gamma^s}}{2\left({\gamma^m}^2-{\gamma^s}^2\right)}f_{\mu\nu}^m\left(f^{\mu\nu}\right)^s\right]
\end{aligned}
\end{eqnarray}
with the following mass-related and spin-related definitions $\lambda_m=\lambda_\uparrow+\lambda_\downarrow$,  $\lambda_s=\lambda_\uparrow-\lambda_\downarrow$, $\theta^m=\left(\theta_\uparrow+\theta_\downarrow\right)/2$, $\theta^s=\left(\theta_\uparrow-\theta_\downarrow\right)/2$, $a^m_\mu=a^\uparrow_\mu+a^\downarrow_\mu$, $a^s_\mu=a^\uparrow_\mu-a^\downarrow_\mu$, $f_{\mu\nu}^m=\partial_\mu a_\nu^m-\partial_\nu a_\mu^m$, and $f_{\mu\nu}^s=\partial_\mu a_\nu^s-\partial_\nu a_\mu^s$.
In terms of mass vortex current $j_{m,\text{vor}}^\lambda=\varepsilon^{\lambda\mu\nu}\partial_\mu\partial_\nu \theta^m_{\text{vortex}}$ and spin vortex current $j_{s,\text{vor}}^\lambda=\varepsilon^{\lambda\mu\nu}\partial_\mu\partial_\nu \theta^s_{\text{vortex}}$, this effective action can be rewritten as 
\begin{eqnarray}
\begin{aligned}\label{EFTChirho}
\mathcal S_{\text{eff}}=\int dtd^2r\sqrt{g}
&\left[ -\frac{{\gamma^m}}{2\left({\gamma^m}^2-{\gamma^s}^2\right)} \left(f_{\mu\nu}^m\right)^2+\ell\,\omega_\mu\varepsilon^{\mu\nu\rho}\partial_\nu a_\rho^m+a_\mu^m j_{m,\text{vor}}^\mu\right.\\
&\,\left.-\frac{{\gamma^m}}{2\left({\gamma^m}^2-{\gamma^s}^2\right)} \left(f_{\mu\nu}^s\right)^2+\mathcal A_{\mu}^{ac}\varepsilon^{\mu\nu\rho}\partial_\nu a_\rho^s+a_\mu^s j_{s,\text{vor}}^\mu+\frac{{\gamma^s}}{2\left({\gamma^m}^2-{\gamma^s}^2\right)}f_{\mu\nu}^m\left(f^s\right)^{\mu\nu}\right]
\end{aligned}
\end{eqnarray}
where the last term represents the interaction between mass gauge field $a^m$ and spin gauge field $a^s$. We emphasize that the last term is important because it embodies curvature-induced spin current and electromagnetic signatures. Note that if spin-up component and spin-down component have different chiralities (e.g. $\rm ^3He$ planar phase), the spin connection will couple to spin current, and one can obtain the term $\left(\ell\,\omega_\mu+\mathcal A_{\mu}^{ac}\right)\varepsilon^{\mu\nu\rho}\partial_\nu a_\rho^s$.
}

While this effective action also contains a Wen-Zee-like term (as was noticed previously \cite{MorozHoyos}), there are two essential differences. First, the gauge field $a_\mu$ is described by the Maxwell-like action $f^{\mu\nu}f_{\mu\nu}$ instead of a Chern-Simons-like action. Therefore, integrating out the auxiliary gauge field leads to very different physical responses. Second, the action contains an Aharonov-Casher gauge field $\mathcal A_\mu^{ac}$ instead of an electromagnetic vector potential $A_\mu$. This leads to a very different electromagnetic response, which we have demonstrated in the main text.

\begin{center}
\textbf{S-IV. Conformal transformation and the propagator of $a_\mu$ on a curved 2D surface.}
\end{center}

To determine the propagator of a gauge field on a curved surface, it is often convenient to employ a conformal transformation. Under a conformal transformation, a metric of a curve surface can be written as 
\begin{equation}
g_{ij}(\bold r)=e^{s(\bold r)}\delta_{ij},
\end{equation}
which differs from the flat space one only by a conformal factor $e^s(\bold r)$. The conformal factor includes all the information of the curved surface, and the propagator of a gauge field on a curved surface will look much neater. To demonstrate the whole mathematical procedure, let us assume a curved surface with finite Gaussian curvature described by a three-dimensional vector $\bold R(r,\varphi)=\left(r\cos \varphi, r\sin\varphi,h_0 \exp{\left(-r^2/2r_0^2\right)}\right)$, where $r$ and $\varphi$ are plane polar coordinates. According to differential geometry, the metric is defined as 
$g_{\mu\nu}=\bold t_\mu\cdot \bold t_\nu$
with $\bold t_\mu\equiv  \partial_\mu \bold R$. Therefore, in polar coordinates, the metric $g_{rr}=\bold t_r\cdot\bold t_r=1+\frac{\alpha^2r^2}{r_0^2}e^{-r^2/r_0^2}$ with $\alpha=h_0/r_0$, $g_{\phi\phi}=1$, and $g_{r\phi}=g_{\phi r}=0$.
Therefore, one can express the area element in polar coordinates, i.e.,
$$ds^2=\left(1+\frac{\alpha^2 r^2}{r_0^2}e^{-\frac{r^2}{r_0^2}}\right)dr^2+r^2 d\phi^2.$$
One could stretch the radial part (i.e., $r\rightarrow \mathcal R(r)$) and assume the new metric to be of the form
$$ds^2=e^{s(\bold r)}\left(d\mathcal R^2 +\mathcal R^2 d\phi^2\right).$$
To ensure the above two metrics describe the same geometry, we require 
\begin{equation}
\frac{d\mathcal R}{\mathcal R}=\sqrt{c(r)}\,\frac{dr}{r}~~~~~~ e^{s(r)}=\frac{r^2}{\mathcal R^2}\nonumber
\end{equation}
with $c(r)\equiv 1+\frac{\alpha^2 r^2}{r_0^2}e^{-\frac{r^2}{r_0^2}}$. The solution reads 
\begin{equation}
\mathcal R=r e^{-\int_r^\infty\frac{d r^\prime}{r^\prime}\left(\sqrt{c(r^\prime)}-1\right)}
\end{equation}
and 
\begin{equation}
s(r)=2\int_r^\infty\frac{d r^\prime}{r^\prime}\left(\sqrt{c(r^\prime)}-1\right).
\end{equation}
This particular solution leaves the origin and the point at the infinity invariant.
The propagator $\Gamma(\bold r, \bold r^\prime)$ can be obtained from solving the following Laplacian equation on a curved surface
\begin{equation}
D^iD_i\Gamma(\bold r, \bold r^\prime)=-\frac{\delta(\bold r, \bold r^\prime)}{\sqrt{g}}
\end{equation}
where $D^iD_i\equiv \left(1/\sqrt{g}\right)\partial_i\left(\sqrt{g}g^{ij}\partial_j\right)$. Under the conformal transformation $\sqrt{g(r, \phi)}\rightarrow e^{s(r)}\sqrt{g(\mathcal R, \phi)}$ and $g^{ij}(r,\phi)\rightarrow e^{-s(r)}g^{ij}(\mathcal R, \phi)$. Then the exponential factors cancel out, leading to a Laplacian of a flat plane in terms of coordinates $\left(\mathcal R(r), \phi\right)$. As a result, the propagator reads
\begin{equation}
\Gamma(\bold r, \bold r^\prime)=-\frac{1}{4\pi}\ln \left[\mathcal R(r)^2+\mathcal R(r^\prime)^2-2 \mathcal R(r)\mathcal R(r^\prime)\cos\left(\phi-\phi^\prime\right)\right]
\end{equation}
Note that $\Gamma(\bold r, \bold r^\prime)$ differs from the real flat space expression only be a stretch of the radial part. 
The geometric potential is defined as 
\begin{equation}\label{apvgeo}
V_{\text{geo}}=\int d^2 r^\prime\sqrt{g}\mathcal B(\bold r^\prime) \Gamma(\bold r^\prime, \bold r).
\end{equation}
Act on both sides of the above equation with the covariant Laplacian and using the definition of the propagator, we can obtain
\begin{equation}\label{didiv1}
D_iD^i V_{\text{geo}}(\bold r)=\mathcal B(\bold r).
\end{equation}
The Gaussian curvature can be expressed in the conformal coordinates: 
\begin{equation}\label{bxy}
\mathcal B(x,y)=-e^{-s(r)}\left(\partial_x^2+\partial_y^2\right)\frac{s(x,y)}{2},
\end{equation} and so does $D_iD^i V_{\text{geo}}(\bold r)$:
\begin{equation}\label{didiv2}
D_i D^iV_{\text{geo}}(\bold r)=e^{-s(x,y)}\left(\partial_x^2+\partial_y^2\right)V_{\text{geo}}(x,y).
\end{equation}
Substitute Eqs.\eqref{bxy} and \eqref{didiv2} into Eq.\eqref{didiv1}, we identify the geometric potential as
\begin{equation}
V_{\text{geo}}(r)=\frac{s(x,y)}{2}=\int_r^\infty d r^\prime \frac{\sqrt{c(r^\prime)}-1}{r^\prime}.
\end{equation}
Thus we have proved the Eq.(14) in the main text by adding back the coefficient $\hbar^2 \rho_s/m$, which we have dropped in writing the propagator. 
In the case of two localized bumps on a surface, the total geometric potential adds up via the formula Eq.\eqref{apvgeo}.

{
\begin{center}
\textbf{S-V. Vortex density and Gaussian curvature attachment.}
\end{center}

In a series of delicate works (see Refs.[24a,24b,29a,29b] in the main text), the authors derived the vorticity and Gaussian curvature attachment by minimizing the following Lagrangian density for chiral superfluids
\begin{eqnarray}
\mathcal L_{\text{ch}}=\frac{J_i J^i}{2\rho}-E(\rho)-\ell\, \omega_\mu J^\mu,
\end{eqnarray}
where $J_\mu$ represents supercurrent density, $E(\rho)$ is the internal energy density, and $\ell$ is the orbital spin. By minimizing this Lagrangian, one can obtain the vorticity equation for charge-neutral superfluids: 
\begin{eqnarray}\label{vortexfluxattach}
j_{\text{vor}}^0=\ell\,\mathcal B,
\end{eqnarray}
where $j_{\text{vor}}^0=\frac{\varepsilon^{ij}\partial_i J_j}{\rho}$ represent vorticity density, and $\mathcal B$ is Gaussian curvature. 

In our formulation, the vorticity and Gaussian curvature attachment is automatically embodied in the geometric potential $V_{\text{geo}}$.  Integrating out $a_0$ in Eq.(12) in the main text yielding a Coulomb gas model with the vortex-geometry interaction
$\ell \int dS\, j^0_{\text{vor}}(\bold r) V_{\text{geo}}(\bold r)$. It should be obvious that vortex density couples to geometric potential in the same manner as electric charge density couples to electric potential. Therefore, a vortex sees an effective electric field $E_i=D_iV_{\text{geo}}(\bold r)$ generated by geometric potential. And according to Gauss law, the total vortex charge inside a region equals the total effective electric flux that goes through the boundary
\begin{eqnarray}
q_{\text{vor}}=\ell\oint_C d l_i D_iV_{\text{geo}}(\bold r)=\ell \int dS D_iD^i V_{\text{geo}}(\bold r)=\ell \int dS \mathcal B(\bold r),
\end{eqnarray}
where we have used Eq.\eqref{didiv1} in the last step.
The differential form of the above equation is consistent with Eq.\eqref{vortexfluxattach} for charge-neutral superfluids.

}

{
\begin{center}
\textbf{S-VI. Self-energy induced anomalous geometric interactions.}
\end{center}

In the main text and above derivations, we have viewed a vortex as a point defect and ignored its self-energy. When considering the self-energy of a vortex, an anomalous geometric interaction emerges, as was nicely reviewed in the literature. This anomalous interaction entails a repulsion of vortices of either sign from positive curvature and an attraction to negative curvature. By sharp contrast, the sign of the geometric interaction discussed in the main text does depend on the sign of vortices: the negative vortices tend to position themselves on peaks and valleys curvature while the positive ones on the saddles of a surface.

One may understand the anomalous geometric interaction by analyzing the streamlines (flow pattern) of a vortex. Streamlines are tangent to the direction of flow, and their density distribution determines the speed of the flow. The number of streamlines becomes sparse on a positive curvature. According to fluid dynamics, sparse streamlines indicate low flowing velocity, and low velocity indicates high pressure. Therefore, vortices are pushed away from positive curvature, regardless of the sign of vortex. Another way to understand this anomalous geometric interaction is by using the electrostatic analogy. The curvature distorts streamlines in a way similar to that a metallic boundary distorts the electric field lines of a charge. According to the image charge method, the interaction is always repulsive between a charge (either positive or negative) and a metallic boundary. Likewise, the interaction of a vortex with curvature produces a force whose sign is independent of the vortex's sense of rotation. 

Beyond understanding the physical picture qualitatively, the conformal mapping technique (discussed in section S-IV) also allows one to calculate the anomalous geometric interaction quantitatively. Let us briefly outline the derivation for normal superfluid vortices\cite{TurnerVitelliNelson}. Let us introduce polar coordinate $r,\varphi$ centered on a vortex of charge $N_{\text{vor}}$. Near the vortex, the energy density is given by $\frac{\gamma (N_{\text{vor}})^2}{8\pi^2 r^2}$. According to the conformal mapping, curvature can either stretch or contract the core radius of a vortex from $a$ to $a^{\prime}=e^{V_{\text{geo}}(\bold r)}a$. On a positive curvature $V_{\text{geo}}>0$, the core radius is stretched; whereas on a negative curvature, the core radius is contracted. Therefore, to calculate the change of the vortex self-energy, one needs to integrate the vortex energy density over the annulus comprised between $a$ and $a^\prime$
\begin{eqnarray}
\delta E_{\text{self}}=\int_{a}^{a^\prime} 2\pi r dr \frac{\gamma  \left(N_{\text{vor}}\right)^2}{8\pi^2 r^2}=\frac{(N_{\text{vor}})^2}{4\pi}V_{\text{geo}},
\end{eqnarray}
which is in sharp contrast to the geometric interaction in the main text $N_{\text{vor}} V_{\text{geo}}$. Including this anomalous geometric interaction yields a total geometric force $N_{\text{vor}}\left(1 + \frac{N_{\text{vor}}}{4\pi} \right) \boldsymbol \nabla V_{\text{geo}}$. Because a vortex's self-energy is proportional to the square of its charge, a vortex of unit charge is therefore energetically favorable. And the anomalous geometric interaction is always smaller than what we considered in the main text.

}

\begin{center}
\textbf{S-VII. Geometric induction in $^3$He planar phase.}
\end{center}

Here we study the anomalous mass current density and spin current density in $\rm ^3$He planar phase, the order parameter of which looks similar to Eq.\eqref{ophea}, but with a key difference \cite{Vollhardtappend}: The spin-up and spin-down component in planar phase possess opposite chiralities
\begin{equation}\label{ophea}
\Psi_P=
\sqrt{\frac{\rho_\uparrow}{2}}\left(\mathbf{\hat e_x}+i\,\mathbf{\hat e_y}\right)e^{i\theta_\uparrow}|\uparrow\rangle+\sqrt{\frac{\rho_\downarrow}{2}}\left(\mathbf{\hat e_x}-i\,\mathbf{\hat e_y}\right)e^{i\theta_\downarrow}|\downarrow\rangle.
\end{equation}
The corresponding GL Lagrangian density looks similar to Eq.(16) in the main text, but the difference is that the spin-up and spin-down component have opposite chirality 
\begin{eqnarray}
\mathcal L_P=\frac{\gamma_\uparrow}{2}\left(\partial_\mu \theta_\uparrow+{\omega_\mu}+\mathcal A^{ac}_\mu\right)^2+\frac{\gamma_\downarrow}{2}\left(\partial_\mu \theta_\downarrow-{\omega_\mu}-\mathcal A^{ac}_\mu\right)^2
\end{eqnarray}  
Following the same logic, we can obtain the anomalous mass current and spin current density driven by space curvature and AC gauge field:
\begin{eqnarray}\label{eqmatrixformp}
\left(
\begin{array}{cc}
j_\mu^{\text{m}}\\  j_\mu^{\text{s}}
\end{array}
\right)
=\left(
\begin{array}{cc}
{\gamma^m} &{\gamma^s}\\
{\gamma^s}&{\gamma^m}
\end{array}
\right)\cdot
\left(
\begin{array}{cc}
\partial_\mu\theta^{\text{m}}\\\omega_\mu+{\mathcal A^{ac}_\mu}
\end{array}
\right),
\end{eqnarray}
where $\partial_\mu\theta^\text{m}=\partial_\mu\theta_\uparrow+\partial_\mu\theta_\downarrow$ is the total phase gradient. Notice that, in contrast to $\rm ^3He$-A phase, the planar phase can in principle preserve time-reversal symmetry, implying $\theta^m=0$, $\rho_s=0$. Therefore, the above formula indicates that there is no mass current, but both the curvature and AC phase can drive spin current.

By minimizing the action, one can also obtain the effective electric charge density and electric current density for the planar phase. The expressions have the exact form as Eq.(18) in the main text except that we need to replace ${\gamma^s}$ with ${\gamma^m}$ since  the spin-up and spin-down components in the planar phase have the same chirality, and therefore the electric currents add up.

\begin{center}
\textbf{S-VIII. Minimization of the GL action for chiral superfluid}
\end{center}

Let us take $^3$He-A (A$_1$ A$_2$) phase as an example and derive the electromagnetic response by minimizing GL action with respective to $\{\phi,  \bold A\}$. 

\textbf{i) Minimized GL action with respect to electric potential $\phi$.---}
We can substitute electric field and magnetic field with gauge potential, i.e. $\bold E=-\boldsymbol\nabla\phi-\partial_t\bold A$ and $\bold B=\boldsymbol\nabla\times\bold A$. 
{In our system, chiral superfluids flow in 2D, whereas electromagnetic field lives in 3D and magnetic moment $\boldsymbol \mu$ can point to any direction in 3D. Therefore, we use 3D vector notations to accommodate electromagnetic fields and the magnetic moment in our system. For example, in 2D electron gas, magnetic field is a scalar $B=\varepsilon^{ij}\partial_i A_j$, which physically represents a magnetic field vector $\bold B=\varepsilon^{ij3}\partial_i A_j \bold{\hat e_3}$ perpendicular to the surface. Similarly, the magnetic-like geometric field strength (i.e., Gaussian curvature) $\mathcal B$ is a scalar field, which corresponds to a vector $\boldsymbol{\mathcal B}=\boldsymbol\nabla\times\boldsymbol{\omega}$ that is perpendicular to the surface. With this preparation, we can express the free energy in terms of vector notations. We first minimize the Lagrangian with respect to electric potential $\phi$, yielding}
\begin{eqnarray}\label{GLphi}
\begin{aligned}
\delta \mathcal L_A&=\gamma_\uparrow\left[\boldsymbol\nabla \theta_\uparrow+\boldsymbol{\omega}+\left(\bold E\times\boldsymbol \mu\right)\right]\cdot \delta(\boldsymbol\nabla\phi\times\boldsymbol \mu)
-\gamma_\downarrow\left[\boldsymbol\nabla\theta_\downarrow+\boldsymbol{\omega}-(\bold E\times\boldsymbol \mu)\right]\cdot \delta(\boldsymbol\nabla\phi\times\boldsymbol \mu)\\
&={\gamma^m}\left[\boldsymbol\nabla\theta^s+(\bold E\times\boldsymbol \mu)\right]\cdot\delta(\boldsymbol\nabla\phi\times\boldsymbol \mu)+{\gamma^s}\left[\boldsymbol\nabla \theta^m+\boldsymbol{\omega}\right]\cdot \delta(\boldsymbol\nabla\phi\times\boldsymbol \mu)\\
&={\gamma^m} \boldsymbol\nabla\delta\phi\cdot \left\{\boldsymbol \mu\times\left[\boldsymbol\nabla\theta^s+(\bold E\times\boldsymbol \mu)\right]\right\}+{\gamma^s} \boldsymbol\nabla\delta\phi\cdot \left[\boldsymbol \mu\times\left(\boldsymbol\nabla \theta^m+\boldsymbol{\omega}\right)\right]\\
&\quad\quad (\text{ In this step, we integrated out total derivatives.})\\
&=-{\gamma^m} ~\delta\phi ~\boldsymbol\nabla\cdot \left\{\boldsymbol \mu\times\left[\boldsymbol\nabla\theta^s+(\bold E\times\boldsymbol \mu)\right]\right\}-{\gamma^s}~ \delta\phi~ \boldsymbol\nabla\cdot \left[\boldsymbol \mu \times\left(\boldsymbol\nabla \theta^m+\boldsymbol{\omega}\right)\right]\\
&={\gamma^m} ~\delta\phi ~\boldsymbol \mu\cdot\boldsymbol\nabla\times\left[\boldsymbol\nabla\theta^s+(\bold E\times\boldsymbol \mu)\right]+{\gamma^s}~ \delta\phi~ \boldsymbol \mu\cdot\boldsymbol\nabla\times\left[\boldsymbol\nabla \theta^m+\boldsymbol{\omega}\right]\\
&\approx {\gamma^s}~ \boldsymbol \mu\cdot \boldsymbol{\mathcal B} \,\delta\phi
\quad\text{(Assume there is no phase singularity.)}
\end{aligned}
\end{eqnarray}
In the above steps, we have used the definitions of mass stiffness ${\gamma^m}=\gamma_\uparrow+\gamma_\downarrow$, spin stiffness ${\gamma^s}=\gamma_\uparrow-\gamma_\downarrow$, mass-current phase $\theta^m=(\theta_\uparrow+\theta_\downarrow)/2$, and spin-current phase $\theta^s=(\theta_\uparrow-\theta_\downarrow)/2$. {In the last step, we ignored the term ${\gamma^m}\boldsymbol \mu \cdot \boldsymbol \nabla \times(\bold E\times\boldsymbol \mu)$, which is a high order contribution compared to the term ${\gamma^s}\boldsymbol \mu\cdot \boldsymbol{\mathcal B}$. This is because, in the absence of an external electric field, the electric field in the AC term has to be induced by curvature. Furthermore, the term ${\gamma^m}\boldsymbol \mu \cdot \boldsymbol \nabla \times(\bold E\times\boldsymbol \mu)$ is proportional to the square of the magnetic moment, making it even higher order. Indeed, with a typical radius of curvature $R_c=100 \rm \mu m$, we can estimate the ratio between the AC field contribution and the geometric contribution: $\frac{\mu^2 E_{\text{ind}}/R_c}{\mu/R_c^2}=e E_{\text{ind}} R_c/m\approx 10^{-15}$, where $\mu=e/m$ is the magnetic moment, and $E_{\text{ind}}$ is the electric field induced by curvature.} Minimization of the total Lagrangian $\mathcal L_{\rm tot}=\mathcal L_{A}-\frac{1}{4}F_{\mu\nu}F^{\mu\nu}$ with respect to $\phi$ leads to the first GL equation that describes charge density distribution:
\begin{eqnarray}\label{GL1}
\sigma_c=-\frac{\delta \mathcal L_A}{\delta \phi}=-{\gamma^s}\,\mu\,\mathcal B.
\end{eqnarray}
{

An alternative approach to minimize the Lagrangian is by using the 2D partial differential form. To do it, we first assume the magnetic moment $\boldsymbol \mu$ perpendicular to the surface, and we can write it as a scalar. Variation of the Lagrangian with respect to electric potential leads to
\begin{eqnarray}
\begin{aligned}
\delta \mathcal L_A &= \gamma_\uparrow\left(\partial_i\theta_\uparrow+\omega_i+\varepsilon_{ik}E^k\mu\right) \varepsilon^{il}\partial_l\delta \phi\, \mu-\gamma_\downarrow\left(\partial_i\theta_\downarrow+\omega_i-\varepsilon_{ik}E^k\mu\right) \varepsilon^{il}\partial_l\delta \phi\, \mu\\
&=\frac{{\gamma^m}+{\gamma^s}}{2} \left[\partial_i\left(\theta^m+\theta^s\right)+\omega_i+\varepsilon_{ik}E^k\mu\right] \varepsilon^{il}\partial_l\delta \phi\, \mu-\frac{{\gamma^m}-{\gamma^s}}{2} \left[\partial_i\left(\theta^m-\theta^s\right)+\omega_i-\varepsilon_{ik}E^k\mu\right] \varepsilon^{il}\partial_l\delta \phi\, \mu\\
&\approx \delta \phi {\gamma^s} \mu \varepsilon^{il}\partial_l \left(\partial_i \theta^m+\omega_i\right)-\delta \phi {\gamma^m} \mu \varepsilon^{il}\partial_l \left(\partial_i \theta^s+\varepsilon_k^j E^k \mu\right)\\
&\approx {\gamma^s}\,\mu\,\mathcal B\delta \phi.
\end{aligned}
\end{eqnarray}
Minimization of the total Lagrangian leads to the same first GL equation as Eq.\eqref{GL1}. While the two approaches are equivalent in our case, the first one may be illuminating for other mixed-dimensional problems. 
}
\\
\textbf{ii) Minimized GL free energy with respect to vector potential $\bold A$.---}Here we derive the second GL equation.\begin{eqnarray}
\begin{split}
\delta \mathcal L_A&=\gamma_{\uparrow}\left[\dot \theta_\uparrow+\omega_0-\boldsymbol \mu\cdot (\boldsymbol \nabla\times\bold A)\right]\left[-\boldsymbol \mu\cdot (\boldsymbol\nabla\times\delta\bold A)\right]+\gamma_{\downarrow}\left[\dot \theta_\downarrow+\omega_0+\boldsymbol \mu\cdot (\boldsymbol \nabla\times\bold A)\right]\left[\boldsymbol \mu\cdot (\boldsymbol\nabla\times\delta\bold A)\right]\\
&\quad -\gamma_\uparrow\left[\boldsymbol\nabla\theta_\uparrow+\boldsymbol{\omega}+(\bold E\times\boldsymbol \mu)\right]\cdot\left[(-\partial_t\delta\bold A)\times\boldsymbol \mu\right]-\gamma_\downarrow\left[\boldsymbol\nabla\theta_\downarrow+\boldsymbol{\omega}-(\bold E\times\boldsymbol \mu)\right]\cdot\left[(\partial_t\delta\bold A)\times\boldsymbol \mu\right]\\
&=\gamma_\uparrow\delta \bold A\cdot \boldsymbol \mu\times\boldsymbol\nabla\left(\dot \theta_\uparrow+\omega_0-\boldsymbol \mu\cdot\bold B\right)-\gamma_\downarrow\delta \bold A\cdot \boldsymbol \mu\times\boldsymbol\nabla\left(\dot \theta_\downarrow+\omega_0+\boldsymbol \mu\cdot\bold B\right)\\
&\quad -\gamma_\uparrow\delta \bold A\cdot\boldsymbol \mu\times\left[\boldsymbol\nabla \dot\theta_\uparrow+\dot{\boldsymbol{\omega}}+\partial_t(\bold E\times\boldsymbol \mu)\right]+\gamma_\downarrow\delta \bold A\cdot\boldsymbol \mu\times\left[\boldsymbol\nabla \dot\theta_\downarrow+\dot{\boldsymbol{\omega}}-\partial_t(\bold E\times\boldsymbol \mu)\right]\\
&=\frac{{\gamma^m}+{\gamma^s}}{2}\delta \bold A\cdot \boldsymbol \mu\times\boldsymbol\nabla\left(\dot\theta_\uparrow+\omega_0-\boldsymbol \mu\cdot \bold B\right)-\frac{{\gamma^m}-{\gamma^s}}{2}\delta \bold A\cdot\boldsymbol \mu\times\boldsymbol\nabla\left(\dot\theta_\downarrow+\omega_0+\boldsymbol \mu\cdot\bold B\right)\\
&\quad -\frac{({\gamma^m}+{\gamma^s}) }{2}\delta \bold A\cdot\boldsymbol \mu\times\left[\boldsymbol\nabla \dot\theta_\uparrow+\dot{\boldsymbol{\omega}}+\partial_t(\bold E\times\boldsymbol \mu)\right]+\frac{({\gamma^m}-{\gamma^s})}{2}\delta \bold A\cdot\boldsymbol \mu\times\left[\boldsymbol\nabla \dot\theta_\downarrow+\dot{\boldsymbol{\omega}}-\partial_t(\bold E\times\boldsymbol \mu)\right]\\
&={\gamma^m}\delta\bold A\cdot \boldsymbol \mu\times\boldsymbol\nabla\left(\dot\theta^s-\boldsymbol \mu\cdot\bold B\right)+{\gamma^s}\delta\bold A\cdot \boldsymbol \mu\times\boldsymbol\nabla(\dot\theta^m+\omega_0)\\
&\quad-{\gamma^m}\delta\bold A\cdot\boldsymbol \mu\times\left[\boldsymbol\nabla\dot\theta^s+\partial_t(\bold E\times\boldsymbol \mu)\right]-{\gamma^s}\delta\bold A\cdot\boldsymbol \mu\times\left[\boldsymbol\nabla\dot\theta^m+\dot{\boldsymbol{\omega}}\right].
\end{split}
\end{eqnarray}
Minimization of the total Lagrangian $\mathcal L_{\rm tot}$ with respect to $\bold A$ leads to the second GL equation describing current density: 
\begin{eqnarray}
\bold J_c=\delta \mathcal L_A/\delta\bold A= {\gamma^m} \boldsymbol \mu\times\boldsymbol\nabla\left(\dot\theta^s-\boldsymbol \mu\cdot\bold B\right)+{\gamma^s}\boldsymbol \mu\times\boldsymbol\nabla(\dot\theta^m+\omega_0)
-{\gamma^m}\boldsymbol \mu\times\left[\boldsymbol\nabla\dot\theta^s+\partial_t(\bold E\times\boldsymbol \mu)\right]-{\gamma^s}\boldsymbol \mu\times\left[\boldsymbol\nabla\dot\theta^m+\dot{\boldsymbol{\omega}}\right]\nonumber
\end{eqnarray}
To the first order approximation in the absence of electromagnetic field, the effective charge current is
\begin{eqnarray}
\bold J_c={\gamma^s}\boldsymbol \mu\times\boldsymbol\nabla\omega_0-{\gamma^s}\boldsymbol \mu\times\dot{\boldsymbol{\omega}}={\gamma^s}\boldsymbol \mu\times \boldsymbol{\mathcal E}.
\end{eqnarray}
given the time-independent superfluid phase. In the above formula $\boldsymbol{\mathcal E}=\left(\mathcal E_1, \mathcal E_2\right)$ where $\mathcal E_i\equiv \epsilon_{0ij}\mathcal E^j$.

{
\begin{center}
\textbf{S-IX. Introducing spin-spin interactions.}
\end{center}

One may further introduce a small spin-spin interaction into the free energy \cite{Leggett1968app}. Without spin-flipping terms, spin is still a good quantum number, and the simplest spin-spin interaction takes the form
\begin{equation}
b\left( \partial_\mu \theta_\uparrow+\omega_\mu+A_\mu^{ac}\right)\left( \partial_\mu \theta_\downarrow+\omega_\mu-A_\mu^{ac}\right),
\end{equation}
where $b$ measures the spin-spin interaction strength. In what follows, we show that this spin-spin interaction term does not affect the electromagnetic response of $\rm ^3$He-A ($A_1$, $A_2$) phase, but it can affect the electromagnetic response of $\rm ^3$He planar phase. 

To see how the electromagnetic response changes, we again need to minimize the free energy including the new term. The variation of the free energy with respect to electric potential $\phi$ yields
\begin{eqnarray}
\begin{aligned}
\delta \mathcal L_A&=\gamma_\uparrow\left[\boldsymbol\nabla \theta_\uparrow+\boldsymbol{\omega}+\left(\bold E\times\boldsymbol \mu\right)\right]\cdot \delta(\boldsymbol\nabla\phi\times\boldsymbol \mu)
-\gamma_\downarrow\left[\boldsymbol\nabla\theta_\downarrow+\boldsymbol{\omega}-(\bold E\times\boldsymbol \mu)\right]\cdot \delta(\boldsymbol\nabla\phi\times\boldsymbol \mu)\\
&~~-b\left[\boldsymbol\nabla\theta_\uparrow+\boldsymbol{\omega}+(\bold E\times\boldsymbol \mu)\right]\cdot \delta(\boldsymbol\nabla\phi\times\boldsymbol \mu)+ b\left[\boldsymbol\nabla\theta_\downarrow+\boldsymbol{\omega}-(\bold E\times\boldsymbol \mu)\right]\cdot \delta(\boldsymbol\nabla\phi\times\boldsymbol \mu)\\
&=\left(\gamma_\uparrow-b\right)\left[\boldsymbol\nabla \theta_\uparrow+\boldsymbol{\omega}+\left(\bold E\times\boldsymbol \mu\right)\right]\cdot \delta(\boldsymbol\nabla\phi\times\boldsymbol \mu)
-\left(\gamma_\downarrow-b\right)\left[\boldsymbol\nabla\theta_\downarrow+\boldsymbol{\omega}-(\bold E\times\boldsymbol \mu)\right]\cdot \delta(\boldsymbol\nabla\phi\times\boldsymbol \mu).
\end{aligned}
\end{eqnarray}
One observes that the spin-spin interaction term effectively shifts the stiffness of spin-up and spin-down components $\gamma_\uparrow^\prime=\gamma_\uparrow-b$ and $\gamma_\downarrow^\prime=\gamma_\downarrow-b$.
As a result, the new effective mass stiffness changes to ${\gamma^m}^\prime=\gamma_\uparrow^\prime+\gamma_\downarrow^\prime={\gamma^m}-2b$, while new effective spin stiffness ${\gamma^s}^\prime=\gamma_\uparrow^\prime-\gamma_\downarrow^\prime={\gamma^s}$ remains invariant. In this case, spin current will not change but mass current is reduced. According to Eq.(\ref{GLphi}) and (\ref{GL1}), the electromagnetic signature will not change. The physical reason is as follows:
In a charge-neutral condensate, all electromagnetic signatures come from the spin current. In $\rm ^3$He-A ($A_1$, $A_2$) phase, spin current density is determined by spin stiffness and curvature. And since the spin stiffness does not change, curvature-induced electromagnetic signatures remain the same. However, in $\rm ^3$He planar phase, the electromagnetic signature does change. Because, in $\rm ^3$He planar phase, the spin current density is determined by mass stiffness and curvature. Therefore, a change of mass stiffness yields a change in electromagnetic signature. By minimizing the Lagrangian of the $\rm ^3$He planar-phase, we obtain the new GL equations for curvature induced charge density $\sigma_c=-\left({\gamma^m}-2b\right)\mu\,\mathcal B$,  and electric current density $\bold J_{c}=\left({\gamma^m}-2b\right)\,\boldsymbol \mu \times\boldsymbol{\mathcal E}(\mathbf r)$.

}

\begin{center}
\textbf{Supplemental References}
\end{center}

\end{widetext}

\begin{thebibliography}{}
 
\bibitem{ShapereW}
See the nice review by A. Shapere and F. Wilczek (eds.), {\it Geometric Phases in Physics} (World Scientific, Singapore, 1989), and references therein.

\bibitem{foucault}
M. Berry, The geometric phase, Scientific American {\bf 259}, 46 (1988).

\bibitem{ShapereW2}
A. Shapere and F. Wilczek,  Phy. Rev. Lett. {\bf 58}, 2051 (1987);
American Journal of Physics, {\bf 57}, 514–518 (1989);
Journal of Fluid Mechanics, {\bf 198}, 557–585 (1989).

\bibitem{ParkL}
H. S. Seung and David R. Nelson, Phys. Rev. A {\bf 38}, 1005 (1988); Park, J.-M., and T. C. Lubensky,  Phys. Rev. E {\bf 53}, 2648, (1996); Bowick, M., D. R. Nelson, and A. Travesset, Phys. Rev. E {\bf 69}, 041102, (2004).


\bibitem{ATuner}
V. Vitelli and Ari M. Turner, Phys. Rev. Lett. {\bf 93}, 215301(2004);
L. Giomi and Mark Bowick, Phys. Rev. B {\bf 76}, 054106 (2007);
H. Jiang, G. Huber, R. A. Pelcovits, and T. R. Powers, Phys. Rev. E {\bf 76}, 031908 (2007).

\bibitem{VinitskiiD}
S. I. Vinitskii, V. L. Derbov, V. M. Dubovik, B. L. Markovski, and Y. P. Stepanovskii, Uspekhi (Sov. Phys.) {\bf 33}, 403–428 (1990).

\bibitem{XiaoCN}
D. Xiao, M.-C. Chang, and Q. Niu, Rev. Mod. Phys. {\bf 82}, 1959 (2010);
E. Cohen, H. Larocque, F. Bouchard, F. Nejadsattari, Y. Gefen, and E. Karimi, Nat. Rev. Phys. {\bf 1}, 437 (2019).

\bibitem{QiZ}
M. Z. Hasan and C. L. Kane, Rev. Mod. Phys. {\bf 82}, 3045 (2010); X.-L. Qi and S.-C. Zhang, Rev. Mod. Phys. {\bf 83}, 1057 (2011).

\bibitem{Maeno}
Y. Maeno, H. Hashimoto, K. Yoshida, S. Nishizaki, T. Fujita, J. Bednorz, and F. Lichtenberg, Nature {\bf 372}, 532 (1994); C. Kallin,  Rep. Prog. Phys. {\bf 75}, 042501 (2012).

\bibitem{Kopnin}
N. B. Kopnin and M. M. Salomaa, Phys. Rev. B {\bf 44}, 9667 (1991);
N. Read and D. Green, Phys. Rev. B {\bf 61}, 10267 (2000).

\bibitem{Volovik}
G. E, Volovik, {\it The Universe in a Helium Droplet} (Oxford University Press, 2003).

\bibitem{Volovik1}
G. E. Volovik and L. P. Gor'kov, JETP Lett. {\bf 39}, 674  (1984); Sov. Phys. JETP {\bf 61}, 843 (1985).

\bibitem{JSauls}
V. Braude and E. B. Sonin, Phys. Rev. B {\bf 74}, 064501 (2006); James A. Sauls, Phys. Rev. B {\bf 84}, 214509 (2011).
Wenxing Nie, Wen Huang, and Hong Yao, Phys. Rev. B {\bf 102}, 054502 (2020).

\bibitem{Append}
See supplementary for details including references \onlinecite{MBowick2009,ZHK1989,stone1990}.

\bibitem{MBowick2009}
Mark J. Bowick, Luca Giomi, Adv. Phys. {\bf 58}, 449 (2009);
D. Francesco, P. Mathieu, and D. S\'en\'echal, Conformal Field Theory (Springer, New York, 1997); A. M. Turner, V. Vitelli, and D. R. Nelson, Rev. Mod. Phys. {\bf 82}, 1301 (2010).

\bibitem{ZHK1989}
S. C. Zhang, T. H. Hansson, and S. Kivelson, Phys. Rev. Lett. {\bf 62}, 82 (1989); N. Read, Phys. Rev. Lett. {\bf 62}, 86 (1989); S. M. Girvin and A. H. MacDonald, Phys. Rev. Lett. {\bf 58}, 1252 (1987).

\bibitem{stone1990}
M. Stone, Phys. Rev. B {\bf 42}, 212 (1990); D.-H. Lee and C. L. Kane, Phys. Rev. Lett. {\bf 64}, 1313 (1990).

\bibitem{KvorningH}
T. Kvorning, T. H. Hansson, A. Quelle, and C. M. Smith, Phys. Rev. Lett. {\bf 120}, 217002 (2018).

\bibitem{JiangHW}
Q.-D. Jiang, T. H. Hansson, and F. Wilczek, Phys. Rev. Lett. {\bf 124}, 197001 (2020);
C. Sp\aa nsl\"{a}tt, Phys. Rev. B {\bf 98}, 054508 (2018).

\bibitem{Vollhardt}
D. Vollhardt and P. W\"{o}lfle, {\it The Superfluid Phases of Helium 3} (Taylor and Francis, USA, 1990).

\bibitem{Nayak}
C. Nayak, S. H. Simon, A. Stern, M. Freedman, and S. Das Sarma, Rev. Mod. Phys. {\bf 80}, 1083 (2008);
J. Alicea, Rep. Prog. Phys. {\bf 75}, 076501 (2012);
C. W. J. Beenakker, Annu. Rev. Con. Mat. Phys. {\bf 4}, 113 (2013).

\bibitem{chung}
Suk Bum Chung, Hendrik Bluhm, and Eun-Ah Kim, Phys. Rev. Lett. {\bf 99}, 197002 (2007);
Y. Tsutsumi, T. Kawakami, T. Mizushima, M. Ichioka, and K. Machida, Phys. Rev. Lett. {\bf 101}, 135302 (2008).

\bibitem{Hoyos}
C. Hoyos, S. Moroz, and D. T. Son, Phys. Rev. B {\bf 89}, 174507 (2014);
A. Shitade and T. Kimura, Phys. Rev. B {\bf 90}, 134510 (2014);
S. Moroz, C. Hoyos, and L. Radzihovsky, Phys. Rev. B {\bf 93}, 024521 (2016);
O. Golan, C. Hoyos, and S. Moroz, Phys. Rev. B {\bf 100}, 104512 (2019);
T. Furusawa, K. Fujii, and Yusuke Nishida, Phys. Rev. B {\bf 103}, 064506 (2021).

\bibitem{GolanS}
J. Nissinen, Phys. Rev. Lett. {\bf 124}, 117002 (2020);
O. Golan and Ady Stern, Phys. Rev. B {\bf 98}, 064503 (2018);
Z.-M. Huang, Bo Han, and Michael Stone, Phys. Rev. B {\bf 101}, 125201 (2020).

\bibitem{Zee}
A standard analytical method by assuming vortices as point defects - see the chapter VI.3 for an excellent introduction in  A. Zee, {\it Quantum Field Theory in a Nutshell} (Orient
Longman, Princeton, 2005).

\bibitem{WenZee1992}
X. G. Wen and A. Zee, Phys. Rev. Lett. {\bf 69}, 953 (1992).

\bibitem{StoneRoy2004}
M. Stone and R. Roy, Phys. Rev. B {\bf 69}, 184511 (2004).

\bibitem{GolkarMoroz}
S. Moroz and C. Hoyos, Phys. Rev. B {\bf 91}, 064508 (2015); S. Golkar, M. M. Roberts and Dam Thanh Son, JHEP {\bf 1504}, 110 (2015).


\bibitem{Mmondal}
Mintu Mondal et al., Phys. Rev. Lett. {\bf 107}, 217003 (2011).

\bibitem{IkegamiTK}
P. M. Walmsley and A. I. Golov, Phys. Rev. Lett. {\bf 109}, 215301 (2012); H. Ikegami, Y. Tsutsumi, and K. Kono,  Science {\bf 341}, 59 (2013).

\bibitem{AharonovC}
Y. Aharonov and A. Casher, Phys. Rev. Lett. {\bf 53}, 319 (1984);
H. Mathur and A. Douglas Stone, Phys. Rev. Lett. {\bf 68}, 2964 (1992);
A. V. Balatsky and B. L. Altshuler, Phys. Rev. Lett. {\bf 70}, 1678 (1993).

\bibitem{Shen}
S. Q. Shen, Phys. Rev. Lett. {\bf 95}, 187203 (2005);
Z. Bao, X. C. Xie, and Q.-F. Sun, Nat. Commun. {\bf 4}, 2951 (2013);
Q.-D. Jiang, Z. Bao, Q.-F. Sun, and X.C. Xie, Scientific Reports, {\bf 5}, 11925 (2015).


\bibitem{Leggett1968}
A. J. Leggett, Ann. Phys. {\bf 46}, 76 (1968); K. Roberts, R. Budakian, and M. Stone, Phys. Rev. B {\bf 88}, 094503 (2013).

\bibitem{Nelson}
M. Bowick and L. Giomi, Adv. Phys. {\bf 58}, 449 (2009); {\it Statistical Mechanics of Membranes and Surfaces}, edited by D. R. Nelson, T. Piran, and S. Weinberg (World Scientific, Singapore, 2004).

\bibitem{flxuralGraphene}
Bruno Amorim and Francisco Guinea, Phys. Rev. B {\bf 88}, 115418 (2013).

\bibitem{Rllan}
R. Ilan, A. G. Grushin, and D. I. Pikulin,  Nat. Rev. Phys. {\bf 2}, 29 (2020); A. Cortijo, Y.  Ferreir\'os, K. Landsteiner, and M. A. H. Vozmediano, Phys. Rev. Lett. {\bf 115}, 177202 (2015); Long Liang and Teemu Ojanen, Phys. Rev. Research {\bf 1}, 032006(R) (2019).

\bibitem{HKim}
H.-H. Kim, S. M. Souliou, M. E. Barber, {\it et al.} Science {\bf 362}, 1040 (2018).

\bibitem{KDunnett}
K. Dunnett, A Narayan, N. A. Spaldin, A. Balatsky, Physical Review B {\bf 97}, 144506 (2018).

\end{thebibliography}

\begin{thebibliography}{}

\bibitem{MBowick2009a}
Mark J. Bowick, Luca Giomi, Adv. Phys. {\bf 58}, 449 (2009).

\bibitem{FrancescoMathieua}
See page 56 in D. Francesco, P. Mathieu, and D. S\'en\'echal, Conformal Field Theory (Springer, New York, 1997).

\bibitem{SeungNelson}
H. S. Seung and David R. Nelson, Phys. Rev. A {\bf 38}, 1005 (1988).

\bibitem{Park1996}
Jeong-Man Park and T. C. Lubensky, Phys. Rev. E {\bf 53}, 2648 (1996).

{
\bibitem{ZHK1989app}
S. C. Zhang, T. H. Hansson, and S. Kivelson, Phys. Rev. Lett. {\bf 62}, 82 (1989); N. Read, Phys. Rev. Lett. {\bf 62}, 86 (1989); S. M. Girvin and A. H. MacDonald, Phys. Rev. Lett. {\bf 58}, 1252 (1987).

\bibitem{stone1990app}
M. Stone, Phys. Rev. B {\bf 42}, 212 (1990); D.-H. Lee and C. L. Kane, Phys. Rev. Lett. {\bf 64}, 1313 (1990).

\bibitem{WenZee1992app}
X. G. Wen and A. Zee, Phys. Rev. Lett. {\bf 69}, 953 (1992). 

\bibitem{StoneRoy2004}
M. Stone and R. Roy, Phys. Rev. B {\bf 69}, 184511 (2004). 

\bibitem{MorozHoyos}
S. Moroz and C. Hoyos, Phys. Rev. B {\bf 91}, 064508 (2015); C. Hoyos, S. Moroz, and Dam Thanh Son, Phys. Rev. B {\bf 89}, 174507 (2014);  S. Golkar, M. M. Roberts and Dam Thanh Son, JHEP {\bf 1504}, 110 (2015).

\bibitem{TurnerVitelliNelson}
A. M. Turner, V. Vitelli, and D. R. Nelson, Rev. Mod. Phys. {\bf 82}, 1301 (2010).

\bibitem{Leggett1968app}
A. J. Leggett, Ann. Phys. {\bf 46}, 76 (1968); K. Roberts, R. Budakian, and M. Stone, Phys. Rev. B {\bf 88}, 094503 (2013).
}

\bibitem{Vollhardtappend}
D. Vollhardt and P. W\"{o}lfle, {\it The Superfluid Phases of Helium 3} (Taylor and Francis, USA, 1990).

\end{thebibliography}
\end{document}